\documentclass[12pt,a4paper]{article}
\usepackage{amssymb,amsmath}
\usepackage{amsfonts}
\usepackage{a4}
\begin{document}
 \baselineskip18pt
	
	\title{\bf Polyadic cyclic codes over a non-chain ring $\mathbb{F}_{q}[u,v]/\langle f(u),g(v), uv-vu\rangle$}
	\author{Mokshi Goyal{\footnote {E-mail: mouliaggarwal701@gmail.com}} ~ and ~ Madhu Raka{\footnote {Corresponding author, e-mail: mraka@pu.ac.in}}
		\\ \small{\em Centre for Advanced Study in Mathematics}\\
		\small{\em Panjab University, Chandigarh-160014, INDIA}\\
		\date{}}
	\maketitle
\maketitle
{\abstract{Let $f(u)$ and $g(v)$ be any two polynomials of degree $k$ and $\ell$ respectively ($k$ and $\ell$ are not both $1$),  which split into distinct linear factors over $\mathbb{F}_{q}$. Let $\mathcal{R}=\mathbb{F}_{q}[u,v]/\langle f(u),g(v),uv-vu\rangle$ be a finite commutative non-chain ring. In this paper, we study polyadic codes and  their extensions  over the ring $\mathcal{R}$.  We  give examples of some polyadic
codes which are optimal  with respect to Griesmer type bound for rings. A Gray map is defined from $\mathcal{R}^n \rightarrow \mathbb{F}^{k\ell n}_q$ which preserves  duality. The Gray images of polyadic codes and their extensions over the ring $\mathcal{R}$  lead to construction of self-dual, isodual, self-orthogonal and complementary dual (LCD) codes over $\mathbb{F}_q$. Some examples are also given to illustrate this.\vspace{2mm}
		}
	\section{ Introduction}
$~~~~$ Polyadic cyclic codes or simply called polyadic codes
form an important class of cyclic codes. They have rich algebraic structures for efficient error detection and correction, which explains their preferred role in engineering.  Polyadic  codes generalize quadratic residue codes, duadic codes, triadic codes and $m$-adic residue codes.\vspace{2mm}

Codes over finite rings have been known for several decades, but interest in these codes
increased substantially after a break-through work by Hammons et al.  in 1994, which shows that some well known binary non-linear codes like Kerdock codes and Preparata codes  can be constructed from  linear codes over $\mathbb{Z}_4$. Since then, a lot of research has been done on  cyclic codes, in particular on quadratic residue codes over different types of finite rings such as integer residue rings $\mathbb{Z}_m$, Galois rings $GR(p^s,m)$, chain rings and non-chain rings. Kaya  et al. \cite{Ka1} and Zhang  et al. \cite{Zh} studied quadratic residue codes over a non-chain ring $\mathbb{F}_p+u\mathbb{F}_p$, where $u^2=u$ and $p$ is an odd prime. Bayram and Siap \cite{BS} considered cyclic and constacyclic codes over $\mathbb{F}_{p}[v]/\langle v^p-v\rangle$, where $p$ is a prime.
Kaya et al. \cite{Ka2} studied quadratic residue codes over  $\mathbb{F}_2+u\mathbb{F}_2+u^2\mathbb{F}_2,$ whereas Liu et al. \cite{Li} studied them over non-local ring $\mathbb{F}_p+u\mathbb{F}_p+u^2\mathbb{F}_p$, where $u^3=u$ and $p$ is an odd prime. The authors \cite{RKG} along with Kathuria   extended their results over the ring
	$\mathbb{F}_{p}+u\mathbb{F}_{p}+u^2\mathbb{F}_{p}+u^3\mathbb{F}_{p}$, where $u^4=u$ and $p\equiv 1 ({\rm mod ~} 3)$. In \cite{GR1},  the authors studied  quadratic residue codes  and their extensions over the ring $\mathbb{F}_{p}+u\mathbb{F}_{p}+u^2\mathbb{F}_{p}+\cdots+u^{m-1}\mathbb{F}_{p}$, where  $u^m=u$, $m$ any integer greater than $1$ and $p$ is a prime satisfying $p\equiv 1 ({\rm mod~}(m-1))$.  In \cite{GR2}, the authors considered a more general non-chain ring  $\mathbb{F}_{q}[u]/\langle f(u)\rangle$, where $q$ is a prime power and $f(u)$ is a polynomial of degree $m\ge 2$, which splits into distinct linear factors over $\mathbb{F}_{q}$ and studied  duadic and triadic codes   over it generalizing all the previous results (the condition that $q\equiv 1 ({\rm mod~}(m-1))$ ensures that $u^m-u$ splits into distint linear factors over $\mathbb{F}_{q}$). In another paper \cite{GR3}, the authors have studied  duadic negacyclic codes   over the ring $\mathbb{F}_{q}[u]/\langle f(u)\rangle$.  In \cite{KOS}, Kuruz et al. studied $m$-adic residue codes over $\mathbb{F}_{q}[v]/\langle v^2-v\rangle$. \vspace{2mm}

Recently people have started studying  codes over finite commutative non-chain rings having 2 or more variables.
 Ashraf and Mohammad \cite{AM} studied cyclic codes over $\mathbb{F}_{p}[u,v]/\langle u^2-1,v^3-v, uv-vu\rangle$. They \cite{AM2} also studied skew-cyclic codes over $\mathbb{F}_{q}+u\mathbb{F}_{q}+v\mathbb{F}_{q},$ where $u^2=u, v^2=v, uv=vu=0$. Srinivasulu and Bhaintwal \cite{SB} studied linear codes over $\mathbb{F}_{2}+u\mathbb{F}_{2}+v\mathbb{F}_{2}+uv\mathbb{F}_{2}$, where $u^2=0, v^2=v, uv=vu$, a non-chain extension of $\mathbb{F}_{2}+u\mathbb{F}_{2}$. Yao, Shi and Sol$\acute{e}$ \cite{YSS} studied skew cyclic codes over $\mathbb{F}_{q}+u\mathbb{F}_{q}+v\mathbb{F}_{q}+uv\mathbb{F}_{q}$, where $u^2=u, v^2=v, uv=vu$ and $q$ is a prime power. Islam and Prakash \cite{IP} studied skew cyclic and skew constacyclic codes
over $\mathbb{F}_{q}+u\mathbb{F}_{q}+v\mathbb{F}_{q}+uv\mathbb{F}_{q}$, where $u^2=u, v^2=v$ and $uv=vu$.

In this paper, we study cyclic and polyadic cyclic codes over a more general ring. Let $f(u)$ and $g(v)$ be any two polynomials of degree $k$ and $\ell$ respectively ($k$ and $\ell$ not both $1$),  which split into distinct linear factors over $\mathbb{F}_{q}$. Let $\mathcal{R}=\mathbb{F}_{q}[u,v]/\langle f(u),g(v),uv-vu\rangle$ be a finite commutative non-chain ring. Here we discuss  cyclic codes and their duals over the ring $\mathcal{R}$, define polyadic codes over $\mathcal{R}$ in terms of idempotent generators and study some of their properties. We also give examples of some codes that have optimal parameters with respect to Griesmer type bound for rings. A Gray map is defined from $\mathcal{R}^n \rightarrow \mathbb{F}^{k\ell n}_q$ which preserves linearity and in some special case preserves duality. The Gray images of polyadic codes over the ring $\mathcal{R}$ and their extensions lead to construction of self-dual, isodual, self-orthogonal and complementary dual (LCD) codes over $\mathbb{F}_q$.  In another paper \cite{GR4}, we will consider polyadic negacyclic codes over the ring  $\mathcal{R}$.\vspace{2mm}
	
	The paper is organized as follows: In Section 2, we give some preliminaries including Griesmer type bound for codes over rings, recall polyadic codes of length $n$ over $\mathbb{F}_q$   and give some of their properties.  In Section 3, we study the ring $\mathcal{R} = \mathbb{F}_{q}[u,v]/\langle f(u),g(v),uv-vu\rangle$, cyclic codes over ring $\mathcal{R}$ and define the Gray map $\Phi$ : $\mathcal{R}^n \rightarrow \mathbb{F}^{k\ell n}_q$. In Section 4, we study polyadic  codes  over $\mathcal{R}$, their extensions, their Gray images and discuss Griesmer type bound for these codes. We give some examples to illustrate our theory.

\section{Preliminaries}

	A cyclic code $\mathcal{C}$ of length $n$ over $\mathbb{F}_q$ can be regarded as an ideal of the ring $\mathbb{S}_n = \mathbb{F}_{q}[x]/\langle x^n-1\rangle$. It has a unique generating polynomial $g(x)$ and a unique idempotent generator $e(x)$. The set $\{ i : \alpha^i {\rm ~is~ a ~zero ~of~} g(x)\}$, where $\alpha$ is a primitive $n$th root of unity in some extension field of $\mathbb{F}_q$,  is called the defining set of $\mathcal{C}$. \vspace{2mm}
	
 A polynomial $a(x)=  \sum_i ~a_ix^i\in \mathbb{S}_n$ is called even-like if $a(1)=0$ otherwise it is called odd-like. A code $\mathcal{C}$ is called even-like if all its codewords are even-like otherwise it is called odd-like.\vspace{2mm}\\ For $(a,n)=1$ , $\mu_a : \mathbb{Z}_n \rightarrow \mathbb{Z}_n$ defined as $\mu_a(i)=ai ({\rm mod~} n)$ is called a multiplier, where $\mathbb{Z}_n$ = $\{0,1,2,...,n-1\}$. It is extended on $\mathbb{S}_n$ by defining  $\mu_a(\sum_i f_ix^i)=\sum_if_ix^{\mu_a(i)}$.\vspace{2mm}\\
 For a linear code $\mathcal{C}$  over $\mathbb{F}_q$, the dual code  $\mathcal{C}^\bot$ is defined as $ \mathcal{C}^\bot =\{x\in \mathbb{F}_q^n~ |~ x\cdot y=0 ~ {\rm for ~ all~} y \in \mathcal{C}\}$, where $x\cdot y$ denotes the usual Euclidean inner product. $\mathcal{C}$ is self-dual if $\mathcal{C}=\mathcal{C}^\bot$ and self-orthogonal if $\mathcal{C}\subseteq \mathcal{C}^\bot$. A code $\mathcal{C}$ is called isodual if it is equivalent to its dual $\mathcal{C}^\bot$. A linear code $\mathcal{C}$ whose dual $\mathcal{C}^\bot$ satisfies $\mathcal{C}\cap\mathcal{C}^\bot=\{0\}$ is called a complementary dual (LCD) code. \vspace{2mm}\\
 Let $ \overline{j}(x)=\frac{1}{n}(1+x+x^2+\cdots+x^{n-1}) $. The even weight $[n,n-1,2]$ cyclic code $\mathbb{E}_n$ over $\mathbb{F}_q$ has generating idempotent $1-\overline{j}(x)$, its dual is the repetition code $[n,1,n]$ with generating idempotent $\overline{j}(x)$.\vspace{2mm}\\
 \noindent The following is a well known result,
  see \cite{HP}:\vspace{2mm}
		
\noindent{\bf  Lemma 1:} (i)  Let $\mathcal{C}$  be a cyclic code of length $n$ over a finite field $\mathbb{F}_q$ with defining set $T$. Then the defining set of  $\mu_a(\mathcal{C})$ is $\mu_{a^{-1}}(T)$ and that of $\mathcal{C}^{\perp}$ is $\mathbb{Z}_n -\mu_{-1}(T)$. \vspace{2mm}

\noindent(ii)  Let $\mathcal{C}$ and $\mathcal{D}$ be cyclic codes of length $n$ over a finite field $\mathbb{F}_q$ with defining sets $T_1$ and $T_2$ respectively. Then  $\mathcal{C}\cap \mathcal{D}$ and $\mathcal{C}+ \mathcal{D}$ are cyclic codes with defining sets $T_1\cup T_2$ and $T_1\cap T_2$ respectively.\vspace{2mm}
		
\noindent (iii) Let $\mathcal{C}$ and $\mathcal{D}$ be cyclic codes of length $n$ over  $\mathbb{F}_q$ generated by the idempotents $E_1, E_2$ in $\mathbb{F}_q[x]/\langle x^n-1\rangle$, then $\mathcal{C}\cap \mathcal{D}$ and $\mathcal{C}+ \mathcal{D}$ are generated by the idempotents $E_1E_2$ and $E_1+E_2-E_1E_2$ respectively.\vspace{2mm}

\noindent(iv) Let $\mathcal{C}$  be a cyclic code of length $n$ over $\mathbb{F}_q$ generated by the idempotent $E$,  then  $\mu_a(\mathcal{C})$  is generated by $\mu_a(E)$ and $\mathcal{C}^{\perp}$ is generated by the idempotent $1-E(x^{-1})$.\vspace{2mm}

A linear code  $\mathcal{C}$ over a finite commutative ring  $R$ is an $R$-submodule of $R^n.$ Dual of a linear code over a finite commutative ring  is defined in the same way and results in Lemma 1 (iii) and (iv) also hold true over any finite ring.

\subsection{Griesmer type bound for codes over rings}

Let $R$ be a finite commutative quasi-Frobenious ring. For a linear code $C$ over $R$, the value $k(C)$ is defined as the rank of minimal free $R$-submodules of $R^n$ which contain $C$. Let $R=\underset{\alpha \in \Lambda}\oplus Re_{\alpha},$ where $e_{\alpha}$ are central orthogonal idempotents with, $1_R=\underset{\alpha \in \Lambda}\sum e_{\alpha}$. Then $R_{\alpha}=Re_{\alpha}$ is also a QF ring for each $\alpha \in \Lambda$. Let $J(R)$ denote the Jacobson radical of $R$. If $C$ is a linear code of length $n$ over $R$, then $C_{\alpha}=Ce_{\alpha}$ is a linear code of length $n$ over $R_{\alpha}$. \vspace{2mm}\\The following Griesmer type bound is due to Shiromoto and  Storme \cite[Theorem 2.6]
{SS}. \vspace{2mm}

\noindent {\bf Theorem 1:} Let $R=\underset{{\alpha} \in \Lambda}\oplus_{}Re_{\alpha}$ be a finite quasi-Frobenious ring such that $R_{\alpha}$ is a local ring for all ${\alpha} \in \Lambda$ and let $q_{\alpha}$ be the prime power such that $|R_{\alpha}/J(R_{\alpha})|=q_{\alpha}$ for each $\alpha \in \Lambda$. If $C$ is a linear code of length $n$ over $R$, then $$n \geq {\displaystyle\sum_{i=0}^{k(C)-1} \bigg\lceil \frac{d(C)}{q^i}\bigg\rceil}$$ where $q= \underset{{\alpha} \in \Lambda}\max\{q_{\alpha}\}$, $k(C)= \underset{{\alpha} \in \Lambda}\max \{k(C_{\alpha})\}$ and
$d(C)= \underset{{\alpha} \in \Lambda}\min\{d(C_{\alpha})\}.$\vspace{2mm}

\noindent The code $C$ over $R$ is said to have parameters $[n,k(C),d(C)].$

\subsection{Polyadic cyclic Codes over $\mathbb{F}_q$ }
 Let  $(n,q)=1$ and suppose \begin{equation} \mathbb{Z}_n=  S_1\cup S_2\cup \cdots \cup S_m \cup S_{\infty}, \end{equation} where\\
	(i) $S_1, S_2, \cdots, S_m$ and $S_{\infty}$ are union of $q$-cyclotomic cosets mod $n$,\vspace{1mm}\\
	(ii) $S_1, S_2, \cdots, S_m$ and $S_{\infty}$ are pairwise disjoint, \vspace{1mm}\\
	(iii) there exists a multiplier $ \mu_a$, $(a,n)=1$ such that $\mu_{a}(S_i)= S_{i+1}, {~\rm for}~ 1\leq i \leq m,$ the subscripts are taken modulo $m$ and $\mu_{a}(S_\infty)= S_\infty.$ \vspace{2mm}\\
It is clear that $0\in S_{\infty}$ always. Let $S_{\infty}'=S_{\infty} - \{0\}$.\vspace{2mm}\\
Then codes, for $1\leq i \leq m$,  having $S_i\cup S_{\infty}'$ or $(S_i\cup S_{\infty})^c$ as their defining sets are called odd-like polyadic  codes and the codes having  $(S_i\cup S_{\infty}')^c$ or $S_i\cup S_{\infty}$ as their defining sets are the associated even-like polyadic  codes. Let $\mathbb{D}_i$ denote the odd-like codes having $S_i\cup S_{\infty}'$ as their defining sets; $\mathbb{D}_i'$ denote the odd-like codes having $(S_i\cup S_{\infty})^c$ as their defining sets; $\mathbb{C}_i$ denote the even-like codes having $(S_i\cup S_{\infty}')^c$ as their defining sets; and $\mathbb{C}_i'$ denote the even-like codes having $S_i\cup S_{\infty}$ as their defining sets. \vspace{2mm}\\
In the special case, when $m=2$ and $S_{\infty}=\{0\}$, polyadic codes are duadic codes \cite{Sm}. When $m=3$, polyadic codes are triadic codes as defined by Pless and Rushanan \cite{PR}.
When $n=p$, an odd prime, $m|(p-1)$, $S_{\infty}=\{0\}$, $\mathbb{Z}^*_p= \langle b \rangle,$ $S_1=\{ b^{mr}, 1 \leq r \leq \frac{p-1}{m}\},~ S_i=b^{i-1}S_1$,  then polyadic codes are $m$-adic residue codes as defined by Job \cite{Job}.
A polyadic code of prime length $p$ exists if and only if  $q$ ia an $m$-adic residue mod $p$, see Brualdi and Pless \cite{BP}. When $n$ is a prime power, the conditions for the existence of polyadic codes over $\mathbb{F}_{q}$ were obtained by Sharma et. al \cite{SBR} and for general $n$ see Bakshi et. al \cite{BRS}.\vspace{2mm}\\
Clearly $\mathbb{D}_1, \mathbb{D}_2, \cdots, \mathbb{D}_m$ are equivalent codes; $\mathbb{D}_1', \mathbb{D}_2', \cdots, \mathbb{D}_m'$ are equivalent;  $\mathbb{C}_1, \mathbb{C}_2, \cdots,$ $ \mathbb{C}_m$ are equivalent; and $\mathbb{C}_1', \mathbb{C}_2', \cdots, \mathbb{C}_m'$ are equivalent codes.\vspace{2mm}\\		
 For $1\leq i \leq m$,	let $e_i(x)$  and $e_i'(x)$ be the even-like idempotent generators of even-like polyadic codes $\mathbb{C}_i$ and   $\mathbb{C}_i'$ respectively, $d_i(x)$ and $d_i'(x)$ be odd-like idempotent generators of odd-like polyadic codes $\mathbb{D}_i$ and   $\mathbb{D}_i'$ respectively.\vspace{2mm}

  As the defining set of $\mathbb{C}_1$ is $S_2\cup S_3 \cup \cdots \cup S_m \cup \{0\}$, the defining set of $\mu_a(\mathbb{C}_1)$ is $\mu_{a^{-1}}(S_2\cup S_3\cup \cdots \cup S_m \cup \{0\})= S_1\cup S_2 \cup \cdots \cup S_{m-1} \cup \{0\}$. Therefore $\mu_a(\mathbb{C}_1)= \mathbb{C}_m$ and hence $\mu_a(e_1)= e_m$. Similarly, $\mu_a(e_i)= e_{i-1}$ for $1 \leq i \leq m$ and $\mu_a(d_i)= d_{i-1}$ for $1 \leq i \leq m$. Similar results hold for $e_i'$ and $d_i'$. \vspace{2mm}

Let the set $\{1,2, \cdots,m\}$ be denoted by $A$. Similar to the properties of triadic codes obtained in \cite{GR2}, we have the following results for polyadic codes over $\mathbb{F}_q.$\vspace{2mm}

		\noindent{\bf Proposition 1:} For any subset $\{t_1, t_2, \cdots, t_r\} \subseteq A,$ where $2\leq r \leq m$, we have \vspace{0.5mm}\\
\noindent(i) $\mathbb{C}_1\cap\mathbb{C}_2\cap \cdots \cap \mathbb{C}_m=\mathbb{C}_{t_1}\cap\mathbb{C}_{t_2}\cap \cdots \cap \mathbb{C}_{t_r} $,\vspace{1mm}\\
(ii) $\mathbb{C}_1+\mathbb{C}_2+\cdots +\mathbb{C}_m= \mathbb{E}_n= \langle x-1\rangle=\langle 1-\overline{j}(x)\rangle $, \vspace{1mm}\\
		(iii)   $\mathbb{D}_1+\mathbb{D}_2+ \cdots +\mathbb{D}_m=\mathbb{D}_{t_1}+\mathbb{D}_{t_2}+\cdots+\mathbb{D}_{t_r}$,\vspace{1mm}\\
		(iv)  $\mathbb{D}_1\cap\mathbb{D}_2\cap \cdots \cap\mathbb{D}_m=  \langle\overline{j}(x)\rangle$,\vspace{1mm}\\
         (v) $\mathbb{C}_i+\mathbb{D}_i= \mathbb{S}_n$, $\mathbb{C}_i\cap\mathbb{D}_i= \{0\} {\rm ~for~} 1 \leq i \leq m$, \vspace{1mm}\\
		(vi)  $e_1(x)e_2(x) \cdots e_m(x)=e_{t_1}(x)e_{t_2}(x)\cdots e_{t_r}(x)$, \vspace{1mm}\\
        (vii) $e_1(x)+e_2(x) + \cdots + e_m(x)-(m-1)e_1(x)e_2(x) \cdots e_m(x ) = 1-\overline{j}(x)$, \vspace{1mm}\\
        (viii) $\sum\limits_{i=1}^m d_i(x)-\underset{i<j}{\sum}d_i(x)d_j(x)+\underset{i<j<k}{\sum}d_i(x)d_j(x)d_k(x)-\cdots (-1)^{m-1}\prod\limits_{i=1}^md_i(x)\vspace{0.5mm}\\ =\sum\limits_{i=1}^r d_{t_i}(x)-\underset{t_i<t_j}{\sum}d_{t_i}(x)d_{t_j}(x)+\underset{t_i<t_j<t_k}{\sum}d_{t_i}(x)d_{t_j}(x)d_{t_k}(x)-\cdots (-1)^{r-1}\prod\limits_{i=1}^rd_{t_i}(x)$,\vspace{1mm}\\
		(ix) $d_1(x)d_2(x) \cdots d_m(x)=\overline{j}(x)$ and \vspace{1mm}\\
         (x) $ d_i(x)= 1-e_i(x),~ e_i(x)d_i(x)=0 {\rm ~for~} 1 \leq i \leq m.$\vspace{2mm}

         \noindent   {\bf  Proof:} By Lemma 1 (ii), the defining set of each of $\mathbb{C}_1\cap\mathbb{C}_2\cap \cdots \cap \mathbb{C}_m$, $\mathbb{C}_{t_1}\cap\mathbb{C}_{t_2}\cap \cdots \cap \mathbb{C}_{t_r}$  is $S_1\cup S_2\cup \cdots \cup S_m \cup\{0\}$, hence they are equal. The defining set of $\mathbb{C}_1+\mathbb{C}_2+ \cdots +\mathbb{C}_m$ is $\{0\}$, which is the defining set of even weight code $\mathbb{E}_n$ having generating idempotent $1-\overline{j}(x)$. Again  by Lemma 1(ii), the defining set of each of $\mathbb{D}_1+\mathbb{D}_2+ \cdots +\mathbb{D}_m$, $\mathbb{D}_{t_1}+\mathbb{D}_{t_2}+\cdots+\mathbb{D}_{t_r}$ is $S_{\infty}'$, hence they are all equal. The defining set of $\mathbb{D}_1\cap\mathbb{D}_2\cap \cdots \cap \mathbb{D}_m$ is $\mathbb{Z}_n- \{0\}$, which is the defining set of the repetition  code  having generating idempotent $\overline{j}(x)$. The defining set of  $\mathbb{C}_i \cap \mathbb{D}_i$ is whole of $\mathbb{Z}_n$, hence it is $\{0\}$ in the ring $\mathbb{S}_n= \mathbb{F}_q[x]/\langle x^n-1\rangle$; whereas defining set of  $\mathbb{C}_i+\mathbb{D}_i$ is $\emptyset$, so it is $\mathbb{S}_n =\langle 1 \rangle$. The other results follow by Lemma 1(iii).  $~~~~~~~~~~~~~~~~~~~~~~~~~~~~~~~~~~~~~~~~~~~~~~~~\Box$ \vspace{2mm}

        \noindent{\bf Proposition 2:} For any subset $\{t_1, t_2, \cdots, t_r\} \subseteq A,$ where $2\leq r \leq m$, we have \vspace{1mm}

\noindent(i) $\mathbb{C}_1'\cap\mathbb{C}_2'\cap \cdots \cap \mathbb{C}_m'= \{0\}$, \vspace{1mm}\\
		(ii) $\mathbb{C}_1'+\mathbb{C}_2'+ \cdots + \mathbb{C}_m'=\mathbb{C}_{t_1}'+\mathbb{C}_{t_2}'+\cdots+\mathbb{C}_{t_r}'$,\vspace{1mm}\\
		(iii)   $\mathbb{D}_1'\cap\mathbb{D}_2'\cap \cdots \cap \mathbb{D}_m'=\mathbb{D}_{t_1}'\cap\mathbb{D}_{t_2}' \cdots \cap \mathbb{D}_{t_r}' $,  \vspace{1mm}\\
		(iv)  $\mathbb{D}_1'+\mathbb{D}_2'+ \cdots +\mathbb{D}_m'=\mathbb{S}_n=\langle 1 \rangle,$\vspace{1mm}\\
		(v) $\mathbb{C}_i'+\mathbb{D}_i'= \mathbb{S}_n$, $\mathbb{C}_i'\cap\mathbb{D}_i'= \{0\}$ for $1 \leq i \leq m$,\vspace{1mm}\\
	(vi)	$\sum\limits_{i=1}^m e_i'(x)-\underset{i<j}{\sum}e_i'(x)e_j'(x)+\underset{i<j<k}{\sum}e_i'(x)e_j'(x)e_k'(x)-\cdots (-1)^{m-1}\prod\limits_{i=1}^m e_i'(x)\vspace{0.5mm}\\= \sum\limits_{i=1}^r~e_{t_i}'(x)-\underset{t_i<t_j}{\sum}e_{t_i}'(x)e_{t_j}'(x)+\underset{t_i<t_j<t_k}{\sum}e_{t_i}'(x)e_{t_j}'(x)e_{t_k}'(x)-\cdots (-1)^{r-1}\prod\limits_{i=1}^re_{t_i}'(x)$,\vspace{1mm}\\
		(vii)  $d_1'(x)d_2'(x) \cdots d_m'(x)=d_{t_1}'(x)d_{t_2}'(x)\cdots d_{t_r}'(x)$, \vspace{1mm}\\
        (viii) $d_1'(x)+d_2'(x) + \cdots + d_m'(x)-(m-1)d_1'(x)d_2'(x) \cdots d_m'(x ) = 1$, \vspace{1mm}\\
		(ix) $e_1'(x)e_2'(x) \cdots e_m'(x)=0$, \vspace{1mm}\\
         (x) $ d_i'(x)= 1-e_i'(x),~ e_i'(x)d_i'(x)=0,\vspace{1mm}$\\
         (xi) $ \mathbb{C}_i + \langle \overline{j}(x) \rangle = \mathbb{D}_i'$, ~$ \mathbb{C}_i \cap \langle \overline{j}(x) \rangle = \{0\}$, \vspace{1mm}\\
         (xii) $ \mathbb{C}_i' + \langle \overline{j}(x) \rangle = \mathbb{D}_i$, ~$ \mathbb{C}'_i \cap \langle \overline{j}(x) \rangle = \{0\}$, \vspace{1mm}\\ (xiii) $\mathbb{C}_i\cap\mathbb{C}_i' = \{ 0 \}, \mathbb{C}_i+\mathbb{C}_i' = \langle 1- \overline{j}(x) \rangle, $
         \vspace{1mm}\\
         (xiv) $\mathbb{D}_i\cap\mathbb{D}_i' = \langle \overline{j}(x) \rangle, \mathbb{D}_i+\mathbb{D}_i' = \mathbb{S}_n,$
         \vspace{1mm}\\
         (xv) $e_i+ \overline{j}(x)=d'_i, ~ e_i'+ \overline{j}(x)=d_i, ~ e_i\overline{j}(x)=0,~ e_i' \overline{j}(x)=0$  and \vspace{1mm}\\
         (xvi) $e_ie'_i= 0, e_i+e'_i=1- \overline{j}(x), d_id'_i= \overline{j}(x), d_i+d'_i=1+\overline{j}(x).$ \vspace{2mm}

\noindent   {\bf  Proof:} Statements (i) to (x) are similar to those of (i) to (x) of Proposition 1. For (xi), we note that the defining set of $\langle \overline{j}(x) \rangle$ is $\mathbb{Z}_n-\{0\}$. Therefore the defining set of $ \mathbb{C}_i \cap \langle \overline{j}(x) \rangle $ is $\mathbb{Z}_n$ and defining set  of $ \mathbb{C}_i + \langle \overline{j}(x) \rangle$ is same as that of $\mathbb{D}_i'$. Similarly we have (xii).  The defining set of $\mathbb{C}_i\cap\mathbb{C}_i' $ is $\mathbb{Z}_n $ and that of $\mathbb{C}_i+\mathbb{C}_i' $ is $\{0\}$. The defining set of $\mathbb{D}_i\cap\mathbb{D}_i' $ is $\mathbb{Z}_n-\{0\} $ and that of $\mathbb{D}_i+\mathbb{D}_i' $ is $\emptyset$. Now (xv) and (xvi) follow  by Lemma 1(iii).$~~~~~~~~~~~~~~~~~~~~~~~~~~~~~~~~~~~~~~~~~~~~~~~~~~\Box$ \vspace{2mm}

	\noindent{\bf Proposition 3:} Suppose $S_{\infty}'$ is empty, then for any subset $\{t_1, t_2, \cdots, t_r\} \subseteq A,$ where $2\leq r \leq m$, we have the following additional results:\vspace{2mm}\\
			(i) $\mathbb{C}_1\cap\mathbb{C}_2\cap \cdots \cap \mathbb{C}_m=\mathbb{C}_{t_1}\cap\mathbb{C}_{t_2}\cap \cdots \cap \mathbb{C}_{t_r}=\{0\}$,\\
		(ii)   $\mathbb{D}_1+\mathbb{D}_2+ \cdots +\mathbb{D}_m=\mathbb{D}_{t_1}+\mathbb{D}_{t_2}+\cdots+\mathbb{D}_{t_r}=\mathbb{S}_n$, \\
	(iii) $\mathbb{D}_1'\cap\mathbb{D}_2'\cap \cdots \cap \mathbb{D}_m'=\mathbb{D}_{t_1}'\cap\mathbb{D}_{t_2}' \cdots \cap \mathbb{D}_{t_r}'= \langle\overline{j}(x)\rangle$, \vspace{1mm}\\
(iv) $\mathbb{C}_1'+\mathbb{C}_2'+ \cdots + \mathbb{C}_m'=\mathbb{C}_{t_1}'+\mathbb{C}_{t_2}'+\cdots+\mathbb{C}_{t_r}'=\mathbb{E}_n= \langle 1-\overline{j}(x)\rangle $, \vspace{1mm}\\	
(v)  $e_1(x)e_2(x) \cdots e_m(x)=e_{t_1}(x)e_{t_2}(x)\cdots e_{t_r}=0$, \vspace{1mm}\\
 (vi) $\sum\limits_{i=1}^m d_i(x)-\underset{i<j}{\sum}d_i(x)d_j(x)+\underset{i<j<k}{\sum}d_i(x)d_j(x)d_k(x)-\cdots (-1)^{m-1}\prod\limits_{i=1}^md_i(x)\vspace{0.2mm} \\= \sum\limits_{i=1}^r~d_{t_i}(x)-\underset{t_i<t_j}{\sum}d_{t_i}(x)d_{t_j}(x)+\underset{t_i<t_j<t_k}{\sum}d_{t_i}(x)d_{t_j}(x)d_{t_k}(x)-\cdots (-1)^{r-1}\prod\limits_{i=1}^rd_{t_i}(x)=1$,\vspace{1mm}\\
(vii)  $d_1'(x)d_2'(x) \cdots d_m'(x)=d_{t_1}'(x)d_{t_2}'(x)\cdots d_{t_r}'(x)=\overline{j}(x)$ and \vspace{1mm}\\
(viii)$\sum\limits_{i=1}^m~e_i'(x)-\underset{i<j}{\sum}e_i'(x)e_j'(x)+\underset{i<j<k}{\sum}e_i'(x)e_j'(x)e_k'(x)-\cdots (-1)^{m-1}\prod\limits_{i=1}^me_i'(x)=\vspace{1mm} \sum\limits_{i=1}^r e_{t_i}'(x)-\underset{t_i<t_j}{\sum}e_{t_i}'(x)e_{t_j}'(x)+\underset{t_i<t_j<t_k}{\sum}e_{t_i}'(x)e_{t_j}'(x)e_{t_k}'(x)-\cdots (-1)^{r-1}
\prod\limits_{i=1}^re_{t_i}'(x)= 1-\overline{j}(x)$.\vspace{1mm}

\noindent Proof is straightforward.\vspace{1mm}\\
			\noindent{\bf Proposition 4:} Let $\mathbb{C}_i$, $\mathbb{C}_i'$, for $1 \leq i \leq m$, be two pairs of even-like polyadic codes over $\mathbb{F}_q$ with $\mathbb{D}_i$, $\mathbb{D}_i'$  the associated pairs of odd-like polyadic codes. Then \vspace{2mm}\\
$~~~~~~~~$ $\mathbb{C}^{\perp}_i=\mu_{-1}(\mathbb{D}_i)$ and  $\mathbb{C}_i'^{\perp}=\mu_{-1}(\mathbb{D}_i')$. \vspace{2mm}\\
	   Further  if $\mu_{-1}(\mathbb{D}_i)=\mathbb{D}_i$,  then \vspace{1mm}\\
$~~~~~~~~$ $\mathbb{C}^{\perp}_i=\mathbb{D}_i$, $\mathbb{C}_i'^{\perp}= \mathbb{D}_i'$ and so $\mathbb{C}_i$, $\mathbb{C}_i'$, $\mathbb{D}_i$ and $\mathbb{D}_i'$ are LCD codes.\vspace{2mm}\\
	\noindent   {\bf  Proof:} As the defining set of  $\mathbb{C}_1$ is $(S_1\cup S_{\infty}')^c= \{0\}\cup S_2\cup S_3 \cup \cdots \cup S_m$, the defining set of $\mathbb{C}^{\perp}_1$, by Lemma 1(i) is \\ $\begin{array}{lll}&= &\mathbb{Z}_n -\mu_{-1}(\{0\}\cup S_2\cup S_3 \cup \cdots \cup S_m)\\&=& \mu_{-1}(\mathbb{Z}_n)- \mu_{-1}(\{0\}\cup S_2\cup S_3 \cup \cdots \cup S_m)\\& =& \mu_{-1}(S_1\cup S_2\cup \cdots \cup S_m\cup X_{\infty})-\mu_{-1}(\{0\}\cup S_2\cup S_3 \cup \cdots \cup S_m)\\&=& \mu_{-1}(S_1\cup X_{\infty}')\\&=& \mu_{-1}( {\rm defining ~ set ~ of ~} \mathbb{D}_1).\end{array}$ \vspace{2mm}\\
This proves that $\mathbb{C}^{\perp}_1=\mu_{-1}(\mathbb{D}_1)$. Similar is the proof of others. When $\mu_{-1}(\mathbb{D}_i)=\mathbb{D}_i$,
we get, from Propositions 1(v) and 2(v), that $\mathbb{C}_i \cap \mathbb{C}^{\perp}_i = \mathbb{C}_i \cap \mathbb{D}_i =\{0\} $ and $\mathbb{C}'_i \cap \mathbb{C}'^{\perp}_i = \mathbb{C}'_i \cap \mathbb{D}'_i =\{0\} $; proving that $\mathbb{C}_i$ and $\mathbb{C}_i'$ are LCD codes. One can check that $\mathbb{D}_i$ and $\mathbb{D}_i'$ are also LCD codes.
$~~~~~~~~~~~~~~~~~~~~~~~~~~~~~~~~~~~~~~~~~~~~~~~~~~~~~~~~~~~~~~~~~~~~~~~~~\Box$

		\section{Cyclic codes over the ring $\mathcal{R}$ and the Gray map}
\subsection{The ring $\mathcal{R}$ }
		Let $q$ be a prime power, $q = p^{s}$. Throughout the paper, $\mathcal{R}$ denotes the commutative ring $\mathbb{F}_{q}[u,v]/\langle f(u),g(v), uv-vu\rangle$, where $f(u)$ and $g(v)$ are polynomials of degree $k$ and $\ell$ respectively, which split into distinct linear factors over $\mathbb{F}_q$. We assume that $k$ and $\ell$ are not both $1$, otherwise $\mathcal{R}\simeq \mathbb{F}_q$. If $\ell=1$ or $k=1$, then the ring $\mathcal{R} = \mathbb{F}_{q}[u,v]/\langle f(u),g(v), uv-vu\rangle$ is isomorphic to $\mathbb{F}_{q}[u]/\langle f(u)\rangle$ or $\mathbb{F}_{q}[v]/\langle g(v)\rangle$. Cyclic, duadic and triadic codes over $\mathbb{F}_{q}[u]/\langle f(u)\rangle$ have  been discussed by the authors in \cite{GR2}. \vspace{2mm}
\\ Let $f(u)=(u-{\alpha}_1)(u-{\alpha}_2)...(u-{\alpha}_k)$, with $\alpha_i \in \mathbb{F}_q$, $\alpha_i \neq \alpha_j$ and $g(v)=(v-{\beta}_1)(v-{\beta}_2)...(v-{\beta}_\ell)$, with $\beta_i \in \mathbb{F}_q$, $\beta_i \neq \beta_j$. $\mathcal{R}$ is a  non chain ring of size ${q}^{k\ell}$ and characteristic $p$. \vspace{2mm}\\
		For $k \geq 2$ and $\ell \geq 2$, let $\epsilon_i$, $1 \leq i \leq k$ and $\gamma_j$, $1\leq j\leq \ell$,  be elements of the ring $\mathcal{R}$ given by\vspace{2mm}
		\begin{equation} \begin{array}{ll}
			\epsilon_i=\epsilon_i(u)= \frac{(u-\alpha_1)(u-\alpha_2)\cdots(u-\alpha_{i-1})(u-\alpha_{i+1})\cdots(u-\alpha_k)}{(\alpha_i-\alpha_1)(\alpha_i-\alpha_2)\cdots(\alpha_i-\alpha_{i-1})
(\alpha_i-\alpha_{i+1})\cdots(\alpha_i-\alpha_k)} {~~\rm and}\vspace{4mm}\\
		\gamma_j=\gamma_j(v)= \frac{(v-\beta_1)(v-\beta_2)\cdots(v-\beta_{j-1})(v-\beta_{j+1})\cdots(v-\beta_\ell)}{(\beta_j-\beta_1)(\beta_j-\beta_2)\cdots(\beta_j-\beta_{j-1})
(\beta_j-\beta_{j+1})\cdots(\beta_j-\beta_\ell)}.
		
		\end{array}\vspace{2mm}\end{equation}
If $k=1$,  we define $\epsilon_i=1$ and if   $\ell =1$, we take  $\gamma_j=1$.\vspace{2mm}

\noindent	 For $ i= 1,2,\cdots,k, j=1,2,...,\ell $, define $\eta_{ij}$ as follows
		\begin{equation}\eta_{ij}=\eta_{ij}(u,v)= \epsilon_i(u) \gamma_j(v). \end{equation}

\noindent{\bf  Lemma ~2:} We have $\eta_{ij}^2=\eta_{ij}, ~\eta_{ij}\eta_{rs}=0~ {\rm ~for~} 1\leq i, r \leq k, 1\leq j, s \leq \ell, (i,j) \neq (r,s) ~{\rm~ and ~}  \sum_{i,j} \eta_{ij}=1$ in $\mathcal{R}$, i.e., $\eta_{ij}$'s are primitive orthogonal idempotents of the ring $\mathcal{R}$.\vspace{1mm}\\
		{\bf Proof:} Since $\epsilon_i \epsilon_r \equiv 0 ~ ( {\rm mod ~} f(u) ) ~ {\rm ~for} ~i\neq r$ and $\gamma_j\gamma_s \equiv 0 ~ ( {\rm mod ~} g(v) ) ~ {\rm ~for} ~j\neq s,$  $\eta_{ij}\eta_{rs}=\epsilon_i\gamma_j\epsilon_r\gamma_s=0$ in $\mathcal{R}.$ To prove $\eta_{ij}^2=\eta_{ij}$, it is enough to prove that $\eta_{ij}(\eta_{ij}-1)=0$ in $\mathcal{R}$. For that we need to prove $(u-\alpha_r)|\eta_{ij}(u,v)(\eta_{ij}(u,v)-1)$ for all $r$ and $(v-\beta_s)|\eta_{ij}(u,v)(\eta_{ij}(u,v)-1)$ for all $s$. If $r \neq i$, then $\eta_{ij}(\alpha_r,v)= \epsilon_i(\alpha_r)\gamma_j(v) =0$.  If $s\neq j$, then $\eta_{ij}(u,\beta_s)= \epsilon_i(u)\gamma_j(\beta_s) =0$, hence $(u-\alpha_r)|\eta_{ij}(u,v)$, for $r \neq i$ and $(v-\beta_s)|\eta_{ij}(u,v),$ for  $s\neq j$. One can easily check that $(u-\alpha_i)|(\eta_{ij}(u,v)-1) {\rm ~and~}(v-\beta_j)|(\eta_{ij}(u,v)-1)$, so $\eta_{ij}(\eta_{ij}-1)=0$ in $\mathcal{R}$ and hence $\eta_{ij}^2=\eta_{ij}$ in $\mathcal{R}$.\vspace{2mm}

\noindent  Now to prove $ \sum_{i,j} \eta_{ij}=1$ in
		 $\mathcal{R}$, it is sufficient to prove that $\sum_{j=1}^\ell \sum_{i=1}^k \eta_{ij}(u, v)\equiv 1 ({\rm mod} ~ (f(u), g(v)))$. This can be easily checked as  $\sum_{j=1}^\ell \sum_{i=1}^k \eta_{ij}(\alpha_r, v)=1$ and $\sum_{j=1}^\ell \sum_{i=1}^k \eta_{ij}(u,\beta_s)=1$ for all $r$ and $s$, $ r= 1,2,\cdots,k, s=1,2,...,\ell $. $~~~\Box$ \vspace{2mm}

\noindent The decomposition theorem of ring theory tells us that $\mathcal{R}= \underset{i,j}{\bigoplus}~\eta_{ij}\mathcal{R}=\underset{i,j}{\bigoplus}~\eta_{ij}\mathbb{F}_q$. \vspace{2mm}
		
\noindent For a linear code $\mathcal{C }$ of length $n$ over the ring $\mathcal{R}$, let for each pair $(i,j), 1 \leq i \leq k, 1 \leq j \leq \ell$, let \vspace{2mm}

	$\mathcal{C }_{ij}= \{ x_{ij}\in \mathbb{F}_{q}^n : \exists ~x_{rs}  \in \mathbb{F}_{q}^n, (r,s) \neq (i,j),  {\rm ~such ~that~} \underset{r,s}{\bigoplus}~\eta_{rs}x_{rs} \in \mathcal{C }\}.$\vspace{2mm}\\
		Then $\mathcal{C}_{ij}$ are linear codes of length $n$ over $\mathbb{F}_{q}$,  $\mathcal{C}=\underset{i,j}{\bigoplus}~\eta_{ij}\mathcal{C}_{ij}
		$ and $|\mathcal{C }|= \underset{i,j} {\prod}|\mathcal{C }_{ij}|$.
		\vspace{2mm}\\
		The following  is  a simple generalization of Theorem 1 of \cite{GR1}.\vspace{2mm}
		
		\noindent{\bf Theorem 2:} Let $\mathcal{C}=\underset{i,j}{\bigoplus}~\eta_{ij}\mathcal{C}_{ij}
		$ be a linear code of length $n$ over $\mathcal{R}$. Then \vspace{2mm}
		
		\noindent (i)~~ $\mathcal{C}$  is cyclic over $\mathcal{R}$ if and only if $\mathcal{C}_{ij}, ~1 \leq i \leq k, 1 \leq j \leq \ell$ are cyclic over $\mathbb{F}_q$.\vspace{2mm}
		
		\noindent (ii) ~~If $\mathcal{C}_{ij}=\langle  g_{ij} (x)\rangle, ~g_{ij}(x)\in \frac{\mathbb{F}_q[x]}{\langle x^{n}-1\rangle}$, $g_{ij}(x)|(x^n-1)$ then,\\ $~~~~~~~~$  $\mathcal{C}=\langle \eta_{11}g_{11}(x),\cdots,\eta_{1\ell}g_{1\ell}(x),\eta_{21}g_{21}(x),\cdots,\eta_{2\ell}g_{2\ell}(x),\cdots,\eta_{k1}g_{k1}(x),\cdots,\eta_{k\ell}g_{k\ell}(x)\rangle$ \\ $~~~~~~~~=\langle g(x)\rangle$, where $g(x)= \sum_{i}\sum_{j} \eta_{ij}g_{ij}$ and $g(x)|(x^{n}-1)$.\vspace{2mm}
		
		\noindent (iii)~~ Further $|\mathcal{C }|=q^{k\ell n-\sum_{j=1}^{\ell} \sum_{i=1}^{k}deg(g_{ij})}$.\vspace{2mm}
		
		\noindent (iv)~~ Suppose that $g_{ij}(x)h_{ij}(x)=x^n-1,~ 1\leq i\leq k, 1\leq j\leq \ell.$ Let $ h(x)=$ \\ $~~~~~~~~$ $\underset{i,j}{\bigoplus}~\eta_{ij}h_{ij}(x),$ then
		$g(x)h(x)=x^n-1$. \vspace{2mm}
		
		\noindent(v)~~~ $ \mathcal{C}^\perp=\underset{i,j}{\bigoplus}~\eta_{ij}\mathcal{C}_{ij}^\perp.$ \vspace{2mm}
		
		\noindent (vi) ~~$ \mathcal{C}^\perp=\langle h^\perp(x)\rangle,$
		 $ h^\perp(x)=\underset{i,j}{\bigoplus}~\eta_{ij}h_{ij}^\perp(x)$,
		 where $h_{ij}^\perp(x)$ is the reciprocal\\ $~~~~~~~~$
		polynomial of $h_{ij}(x), ~1\leq i\leq k, 1\leq j\leq \ell.$  \vspace{2mm}
		
		\noindent(vii)$~~ |\mathcal{C}^\perp|=q^{\sum_{j=1}^\ell\sum_{i=1}^k deg(g_{ij})}$.		
		
\subsection{ The Gray map}
Every element $r(u,v)$ of the ring $\mathcal{R}=\mathbb{F}_{q}[u,v]/\langle f(u),g(v), uv-vu\rangle$ can be uniquely expressed as $$ r(u,v) = {\displaystyle\bigoplus_{i,j}}~\eta_{ij}a_{ij},$$ where $a_{ij} \in \mathbb{F}_q$ for $1 \leq i \leq k, 1 \leq j \leq \ell$. \vspace{2mm}
 	
\noindent Define a Gray map $\Phi : \mathcal{R}\rightarrow \mathbb{F}_q^{k\ell}$  by $$ r(u,v) = \bigoplus_{i,j} \eta_{ij}a_{ij} \longmapsto(a_{11},a_{12},\cdots, a_{1\ell},a_{21},a_{22},\cdots, a_{2\ell},\cdots,a_{k1},a_{k2},\cdots, a_{k\ell})V$$
	where $V$ is any nonsingular matrix over $\mathbb{F}_q$ of order $k\ell\times k\ell$.
		This map can be extended from $\mathcal{R}^n$ to  $(\mathbb{F}_q^{k\ell})^n$ component wise.\vspace{2mm}
		
		\noindent Let the Gray weight of an element $r \in \mathcal{R}$ be $w_{G}(r) =w_H(\Phi(r))$, the Hamming weight of $\Phi(r)$. The Gray weight of  a codeword
		$c=(c_0,c_1,\cdots,c_{n-1})$ $\in \mathcal{R}^n$ is defined as $w_{G}(c)=\sum_{i=0}^{n-1}w_{G}(c_i)=\sum_{i=0}^{n-1}w_H(\Phi(c_i))=w_H(\Phi(c))$. For any two elements $c_1, c_2 \in \mathcal{R}^n$, the Gray distance $d_{G}$ is given by $d_{G}(c_1,c_2)=w_{G}(c_1-c_2)=w_H(\Phi(c_1)-\Phi(c_2))$. \vspace{2mm}

		\noindent \textbf{Theorem 3:} The Gray map $\Phi$ is an  $\mathbb{F}_q$ - linear, one to one and onto map. It is also distance preserving map from ($\mathcal{R}^n$, Gray distance $d_{G}$) to ($\mathbb{F}_q^{k\ell n}$, Hamming distance $d_H$). Further if the matrix $V$  satisfies $VV^T=\lambda I_{k\ell}$, $\lambda \in \mathbb{F}_q^*$, where $V^T$ denotes the transpose of the matrix $V$, then $\Phi(\mathcal{C}^{\perp})=(\Phi(\mathcal{C}))^{\perp}$ for any linear code   $\mathcal{C}$  over $\mathcal{R}$. \vspace{2mm}\\
		\noindent \textbf{Proof.} The first two assertions hold as $V$ is an invertible matrix over $\mathbb{F}_q$.\\ Let now $V=(V_{1}^T, V_{2}^T, \cdots, V_{k\ell}^T) $, where\vspace{2mm}\\ $V_i=(v_{11}^i,v_{12}^i,\cdots,v_{1\ell}^i,v_{21}^i,v_{22}^i,\cdots,v_{2\ell}^i,\cdots v_{k1}^i,v_{k2}^i,\cdots,v_{k\ell}^i)$ is a $1$x$k\ell$ row vector,\vspace{2mm}\\ satisfying $VV^T=\lambda I_{k\ell}$. So that
		\begin{equation}\begin{array}{l} {\sum\limits_{t=1}^{k\ell}}~ (v_{rs}^t)^2=\lambda ~~{\rm for ~all ~} 1\leq r \leq k, \ 1\leq s \leq \ell {\rm ~~and ~~}  {\sum\limits_{t=1}^{k\ell}}~ v_{rs}^tv_{wy}^t=0 ~~{\rm for ~} (r,s) \neq (w,y). \end{array}\end{equation}
		Let $\mathcal{C}$ be a linear code over $\mathcal{R}$. Let $r=(r_0,r_1,\cdots,r_{n-1})\in \mathcal{C}^{\perp}$,  $s=(s_0,s_1,\cdots,s_{n-1})$
 $\in \mathcal{C}$, where $r_i=\eta_{11}a_{11}^i+\eta_{12}a_{12}^i+\cdots+\eta_{kl} a_{k\ell}^i~$ and $s_i=\eta_{11}b_{11}^i+\eta_{12}b_{12}^i+\cdots+\eta_{kl} b_{k\ell}^i$. So that $r\cdot s=0$. It is enough to prove that $\Phi(r)\cdot \Phi(s)=0 $. Using the properties of $\eta_{ij}$'s from Lemma 2, we get
$$ r_i s_i = \eta_{11}a_{11}^ib_{11}^i+\eta_{12}a_{12}^ib_{12}^i+\cdots+\eta_{k\ell} a_{k\ell}^ib_{k\ell}^i.$$ Then
$$\begin{array}{l}0=r\cdot s=\sum\limits_{i=0}^{n-1}r_is_i= \sum\limits_{i=0}^{n-1}~\sum\limits_{r=1}^{k}~\sum\limits_{s=1}^{\ell}~\eta_{rs}\hspace{0.5mm}a_{rs}^i\hspace{0.5mm}b_{rs}^i=\sum\limits_{r=1}^{k}~\sum\limits_{s=1}^{\ell}~\eta_{rs} \Big( \sum\limits_{i=0}^{n-1}a_{rs}^i\hspace{0.5mm}b_{rs}^i\Big)\end{array} $$  implies that
 \begin{equation}\begin{array}{l}{\sum\limits_{i=0}^{n-1}}a_{rs}^ib_{rs}^i=0, ~~~~~{\rm for ~ all ~}r,s,  1 \leq r \leq k, 1 \leq s \leq \ell.\end{array}\end{equation}
Now
		$$\begin{array}{l}\Phi(r_i)=( a_{11}^i,a_{12}^i,\cdots,a_{k\ell}^i)V =\Big(\sum\limits_{r=1}^{k}\sum\limits_{s=1}^{\ell}~a_{rs}^iv_{rs}^1,\sum\limits_{r=1}^{k}\sum\limits_{s=1}^{\ell}~a_{rs}^iv_{rs}^2~,\cdots, \sum\limits_{r=1}^{k}\sum\limits_{s=1}^{\ell}~a_{rs}^iv_{rs}^{k\ell}\Big)\end{array}$$
Similarly $$\begin{array}{l}\Phi(s_i) =\Big(\sum\limits_{w=1}^{k}\sum\limits_{y=1}^{\ell}~b_{wy}^i\hspace{0.5mm}v_{wy}^1,\sum\limits_{w=1}^{k}\sum\limits_{y=1}^{\ell}~b_{wy}^i\hspace{0.5mm}v_{wy}^2~,\cdots, \sum\limits_{w=1}^{k}\sum\limits_{y=1}^{\ell}~b_{wy}^i\hspace{0.5mm}v_{wy}^{k\ell}\Big).\end{array}$$
 Using (4) and (5), we find that
$$\begin{array}{ll}\Phi(r)\cdot \Phi(s)\hspace{-0.5mm}&= {\sum\limits_{i=0}^{n-1}}\Phi(r_i)\cdot \Phi(s_i)= {\sum\limits_{i=0}^{n-1}}\hspace{0.5mm}  {\sum\limits_{t=1}^{k\ell}}\hspace{0.5mm}{\sum\limits_{w=1}^k}\hspace{0.5mm}{\sum\limits_{y=1}^\ell}\hspace{0.5mm}{\sum\limits_{r=1}^{k}}\hspace{0.5mm}{\sum\limits_{s=1}^{\ell}}a_{rs}^i\hspace{0.5mm}
b_{wy}^i\hspace{0.7mm}v_{rs}^t\hspace{0.5mm}v_{wy}^t\vspace{2mm}\\&
		=\sum\limits_{i=0}^{n-1} {\sum\limits_{\substack{{r=1 }\\ {w=r}}}^{k}}\hspace{0.5mm} {\sum\limits_{\substack{{s=1 }\\ {y=s}}}^{\ell}}a_{rs}^i\hspace{0.5mm}b_{rs}^i\Big(\sum\limits_{t=1}^{k\ell}\hspace{0.5mm}(v_{rs}^t)^2\Big)+ {\sum\limits_{i=0}^{n-1}}{\sum\limits_{w=1}^k}{\sum\limits_{y=1}^\ell}\hspace{-0.5mm}{\sum\limits_{\substack{{r=1 }\\ {(r,s)\neq (w,y)}}}^{k}}\hspace{-0.5mm}{\sum\limits_{s=1}^{\ell}}a_{rs}^i\hspace{0.5mm}
		b_{wy}^i \Big(\sum\limits_{t=1}^{k\ell} v_{rs}^{t}\hspace{0.5mm} v_{wy}^{t}\Big )\vspace{3mm}\\&=\lambda {\sum\limits_{i=0}^{n-1}}~\sum\limits_{r=1}^{k}\sum\limits_{s=1}^{\ell}a_{rs}^i\hspace{0.5mm}b_{rs}^i\vspace{2mm}\\&=\lambda \sum\limits_{r=1}^{k}\sum\limits_{s=1}^{\ell}\Big(\sum\limits_{i=0}^{n-1}a_{rs}^i\hspace{0.5mm}b_{rs}^i\Big)=0,\end{array}$$
		 which proves the result. $~~~~~~~~~~~~~~~~~~~~~~~~~~~~~~~~~~~~~~~~~~~~~~~~~~~~~~~~~~~~~\Box$

		\section{Polyadic Codes over the ring $\mathcal{R}$ }

		\noindent We now define polyadic codes of length $n$ over the ring $\mathcal{R}$ in terms of their idempotent generators with the assumption that the conditions on $n$ and $q$ for existence of polyadic codes over the field $\mathbb{F}_q$ are satisfied. Let $\eta_{ij}, ~1\leq i \leq k, 1\leq j \leq \ell,$ be idempotents as defined in (2) and (3). Let the set of ordered  suffixes $\{ij, 1 \leq i \leq k, 1 \leq j \leq \ell\}$  be divided into $m$ disjoint subsets
\begin{equation} \{ij, 1 \leq i \leq k, 1 \leq j \leq \ell\}=A_1 \cup A_2  \cup \cdots \cup A_m \end{equation}

\noindent with the assumption that  each of the sets $A_i$ is non-empty, if $k\ell \geq m$. In that case let $|A_i|=r_i, 1 \leq r_i \leq k\ell-m+1.$\\  If $k\ell < m$, we assume that in the partition (6), $k\ell$ sets are non-empty, each containing exactly one element and the remaining $m-k\ell$ sets are empty. Therefore
$|A_i|=r_i=1$, if $A_i$ is non empty and $|A_i|=r_i=0$, if $A_i$ is empty.

\noindent Therefore  $$k\ell= r_1+r_2+\cdots+r_m.$$
Define $$\theta_{r_1}={~\displaystyle \sum_{ij\in A_1 } \eta_{ij}},$$

$$\theta_{r_2}={~\displaystyle \sum_{ij\in A_2 } \eta_{ij}},$$
$$\cdots ~~~~~~~~~~~~~\cdots $$
$$\theta_{r_m}={~\displaystyle \sum_{ij\in A_m} \eta_{ij}},$$

\noindent with the convention that empty sum is regarded as zero.

 \noindent Using Lemma 2, we find that  \begin{equation}  \theta_{r_1}+ \theta_{r_2}+\cdots+\theta_{r_m}=1,\end{equation} and that $\theta_{r_i}, 1 \leq i \leq m$ are mutually orthogonal idempotents in the ring $\mathcal{R}$,  i.e., \begin{equation} \theta_{r_i}^2=\theta_{r_i} {~\rm for~ all~~ } i,~ \theta_{r_i}\theta_{r_j}=0, {~\rm for~ all~~ } i \neq j. \end{equation}
For $i=1,2, \cdots, m$, let $e_i,~e'_i, d_i,~d'_i$ be the idempotent generators of polyadic codes over $\mathbb{F}_q$ as defined in Section 2.2.\\
\noindent For each tuple $(r_1,r_2,\cdots,r_m),$ let
 \begin{equation}\begin{array}{ll} F_1=F_1^{(r_1,r_2,\cdots,r_m)}&=  \theta_{r_1}d_1+ \theta_{r_2}d_2+\cdots+\theta_{r_m}d_m \vspace{1mm}\\
 F_2=F_2^{(r_1,r_2,\cdots,r_m)}=\mu_a(F_1)&=  \theta_{r_1}d_m+ \theta_{r_2}d_1+\cdots+\theta_{r_m}d_{m-1} \vspace{1mm}\\
 \cdots ~~~~~~~~~~~~~\cdots~~ &\cdots ~~~~~~~~~~~~~\cdots ~~~~~~~~~~~~\cdots \\
 F_m=F_m^{(r_1,r_2,\cdots,r_m)}=\mu_a(F_{m-1})&=  \theta_{r_1}d_2+ \theta_{r_2}d_3+\cdots+\theta_{r_m}d_1 \vspace{2mm}\\

 F'_1=F_1'^{(r_1,r_2,\cdots,r_m)}&=  \theta_{r_1}d'_1+ \theta_{r_2}d'_2+\cdots+\theta_{r_m}d'_m \vspace{1mm}\\
 F'_2=F_2'^{(r_1,r_2,\cdots,r_m)}=\mu_a(F'_1)&=  \theta_{r_1}d'_m+ \theta_{r_2}d'_1+\cdots+\theta_{r_m}d'_{m-1} \vspace{1mm}\\
 \cdots ~~~~~~~~~~~~~\cdots~~ &\cdots ~~~~~~~~~~~~~\cdots ~~~~~~~~~~~~\cdots \\
 F'_m=F_m'^{(r_1,r_2,\cdots,r_m)}=\mu_a(F'_{m-1})&=  \theta_{r_1}d'_2+ \theta_{r_2}d'_3+\cdots+\theta_{r_m}d'_1
 \end{array}
 \end{equation}
\noindent be odd-like idempotents in the ring $\mathcal{R}[x]/\langle x^n-1\rangle$. Similarly let
\begin{equation}\begin{array}{l} E_1=E_1^{(r_1,r_2,\cdots,r_m)}=  \theta_{r_1}e_1+ \theta_{r_2}e_2+\cdots+\theta_{r_m}e_m \vspace{1mm}\\
E_2=\mu_a(E_1),  E_3=\mu_a(E_2),\cdots , E_m=\mu_a(E_{m-1})\vspace{1mm}\\
E'_1=E_1'^{(r_1,r_2,\cdots,r_m)}=  \theta_{r_1}e'_1+ \theta_{r_2}e'_2+\cdots+\theta_{r_m}e'_m \vspace{1mm}\\

 E'_2=\mu_a(E'_1),  E'_3=\mu_a(E'_2),\cdots ,E'_m=\mu_a(E'_{m-1})
\end{array}
\end{equation}

\noindent be even-like idempotents in the ring $\mathcal{R}[x]/\langle x^n-1\rangle$.

\vspace{2mm}

\noindent For each tuple $(r_1,r_2,\cdots,r_m)$, and for each $i, 1\leq i \leq m$, let $ T_i^{(r_1,r_2,\cdots,r_m)},$ $ T_i'^{(r_1,r_2,\cdots,r_m)}$ denote the odd-like polyadic codes  and $  P_i^{(r_1,r_2,\cdots,r_m)},~  P_i'^{(r_1,r_2,\cdots,r_m)}$ denote the even-like polyadic codes over $\mathcal{R}$ generated by the corresponding  idempotents, i.e.
		
	\begin{equation}\begin{array}{l}T_i^{(r_1,r_2,\cdots,r_m)}= \langle F_i^{(r_1,r_2,\cdots,r_m)}\rangle ,~~
		T_i'^{(r_1,r_2,\cdots,r_m)}= \langle F_i'^{(r_1,r_2,\cdots,r_m)}\rangle ,\vspace{1mm}\\
		 P_i^{(r_1,r_2,\cdots,r_m)}= \langle E_i^{(r_1,r_2,\cdots,r_m)}\rangle ,~~
		P_i'^{(r_1,r_2,\cdots,r_m)}= \langle E_i'^{(r_1,r_2,\cdots,r_m)}\rangle.\end{array}\end{equation}
		\\
 Clearly for any tuple $(r_1,r_2,\cdots,r_m)$,  $T_{1}^{(r_1,r_2,\cdots,r_m)}$,  $T_{2}^{(r_1,r_2,\cdots,r_m)}, \cdots,T_{m}^{(r_1,r_2,\cdots,r_m)}$ are equivalent;   $T_{1}'^{(r_1,r_2,\cdots,r_m)}$,  $T_{2}'^{(r_1,r_2,\cdots,r_m)}, \cdots, T_{m}'^{(r_1,r_2,\cdots,r_m)}$ are equivalent;  \\ $P_{1}^{(r_1,r_2,\cdots,r_m)}$,  $P_{2}^{(r_1,r_2,\cdots,r_m)}, \cdots, P_{3}^{(r_1,r_2,\cdots,r_m)}$ are equivalent and  $P_{1}'^{(r_1,r_2,\cdots,r_m)}$,  $P_{2}'^{(r_1,r_2,\cdots,r_m)}, \cdots,$ $ P_{m}'^{(r_1,r_2,\cdots,r_m)}$ are equivalent.\\

 Next we compute the number of inequivalent odd-like and even-like polyadic codes over the ring $\mathcal{R}$.\vspace{2mm}\\
				\noindent{\bf Theorem 4:} If $k\ell \geq m$, then there are\vspace{4mm}\\
 ${\frac{2}{m}{~ \sum\limits_{r_{m-1}=1}^{k\ell-(r_1+r_2\cdots +r_{m-2})-1}}\cdots{~ \sum\limits_{r_2=1}^{k\ell-r_1-(m-2)}}~ { \sum\limits_{r_1=1}^{k\ell-(m-1)}}\binom{k\ell}{r_1}\binom {k\ell-r_1}{r_2}\cdots \binom {k\ell-(r_1+r_2+\cdots+r_{m-2})}{r_{m-1}}}$\vspace{4mm}\\ inequivalent odd-like polyadic codes and the same number of inequivalent even-like polyadic codes over the ring $\mathcal{R}$.\\ If $k\ell < m,$ then there are $$\frac{2}{m}(k\ell)!\binom {m}{k\ell}$$ inequivalent odd-like polyadic codes and the same number of inequivalent even-like polyadic codes over the ring $\mathcal{R}$.\vspace{2mm}

\noindent {\bf Proof :} Let first $k\ell \geq m$, out of $k\ell$ idempotents $\eta_{ij}, 1\leq i\leq k, 1\leq j\leq \ell$, $\theta_{r_1}$ can be chosen in $\binom{k\ell}{r_1}$ ways. Out of remaining $(k\ell-r_1)$ idempotents   $\theta_{r_2}$ can be chosen in $\binom{k\ell-r_1}{r_2}$ ways, continuing like this $\theta_{r_{m-1}}$ can be chosen in $\binom{k\ell-(r_1+r_2+\cdots+r_{m-2})}{r_{m-1}}$ ways and $\theta_{r_m}$ will be fixed. As each $\theta_{r_i}, 1 \leq i \leq m$ must have at least one $\eta_{ij}$, the number of choices of idempotents $\theta_{r_1}d_1+ \theta_{r_2}d_2+\cdots+\theta_{r_m}d_m $ is \vspace{4mm}\\
 ${ \sum\limits_{r_{m-1}=1}^{k\ell-(r_1+r_2\cdots +r_{m-2})-1}}\cdots{~ \sum\limits_{r_2=1}^{k\ell-r_1-(m-2)}}~ { \sum\limits_{r_1=1}^{k\ell-(m-1)}}\binom{k\ell}{r_1}\binom {k\ell-r_1}{r_2}\cdots \binom {k\ell-(r_1+r_2+\cdots+r_{m-2})}{r_{m-1}}$.\vspace{4mm}\\
Since $\mu_a(F_1)=F_2, \mu_a(F_2)=F_3,\cdots, \mu_a(F_{m})=F_1$, and $F_{i}'^{(r_1,r_2\cdots,r_m))
	}$'s contribute equal number of inequivalent odd-like idempotents, we get the desired number. \\
Let now $k\ell<m$. Firstly the $k\ell$ non-empty sets $A_i$ in the partition (6) can be chosen in $\binom {m}{k\ell}$
ways. Out of
$k\ell$ idempotents $\eta_{ij}, 1\leq i\leq k, 1\leq j\leq \ell$, first non-zero $\theta_{r_i}$ can be chosen in $k\ell$ ways, next non-zero $\theta_{r_j}$ can be chosen in $k\ell-1$ ways, $\cdots,$ so the number of choices of  $F_1= \theta_{r_1}d_1+ \theta_{r_2}d_2+\cdots+\theta_{r_m}d_m $ is $(k\ell)! \binom {m}{k\ell}$. Since $\mu_a(F_1)=F_2, \mu_a(F_2)=F_3,\cdots, \mu_a(F_{m})=F_1$, and $F_{i}'$'s contribute equal number of inequivalent odd-like idempotents, we get the required number.
 $~~~~~~~~~~~~~~~~~~~~~~~~~~~~~~~~~~~~~~~~~~~~~~~~~~~~~~~~~~~~~~~~~~~~~~~~~~~~~~~~~~~~~~~~\Box$\vspace{2mm}\\
\noindent We drop the superscript $(r_1,r_2\cdots,r_m)$, when there is no confusion with the idempotents or the corresponding polyadic codes.\vspace{2mm}
		
		\noindent{\bf Theorem 5:} For any subset $\{t_1, t_2, \cdots, t_r\} \subseteq A=\{1,2, \cdots, m\},$ where $2\leq r \leq m$,
		the following assertions hold for polyadic codes over $\mathcal{R}$.
		\vspace{2mm}\\$\begin{array}{ll}
		{\rm (i)}& T_1\cap T_2\cap \cdots \cap T_m=  \langle \overline{j}(x)\rangle, {\rm~ the ~repitition~ code~ over~ } \mathcal{R}\vspace{2mm}\\
		{\rm (ii)}& T_1+T_2+\cdots +T_m= T_{t_1}+T_{t_2}+\cdots + T_{t_r},  \vspace{2mm}\\
		{\rm (iii)}& P_1\cap P_2\cap \cdots \cap P_m= P_{t_1}\cap P_{t_2}\cdots \cap P_{t_r},\vspace{2mm}\\
		{\rm (iv)}& P_1+ P_2+\cdots+P_m= \langle 1-\overline{j}(x)\rangle, {\rm ~the ~even ~weight~ code~ over~ } \mathcal{R} \vspace{2mm}\\
		{\rm (v)} &P_i\cap \langle \overline{j}(x)\rangle = \{0\}, ~T_i\cap \langle \overline{j}(x)\rangle = \langle\overline{j}(x)\rangle {\rm~and} \vspace{2mm}\\
{\rm (vi)}& P_i+ T_i = \mathcal{R}[x]/\langle x^n-1\rangle, ~P_i\cap T_i = \{0\}.\end{array}$\vspace{2mm}\\
	\noindent \textbf{Proof:}
		From the definitions and relations (7)-(10), we find that the sums of products of terms from $F_1, F_2, \cdots, F_m$ taken one at a time, taken two at a time and so on is equal to the sums of products of terms from $d_1, d_2, \cdots, d_m$ taken one at a time, taken two at a time and so on, i.e.
\begin{equation}\begin{array}{l} F_1+F_2+\cdots+F_m=d_1+d_2+\cdots+d_m, \vspace{2mm}\\
\sum\limits_{\substack{{i,j}\\{i<j}}}F_i(x)F_j(x)=\sum\limits_{\substack{{i,j}\\{i<j}}}d_i(x)d_j(x),\vspace{2mm}\\ \sum\limits_{\substack{{i,j,k}\\{i<j<k}}}F_i(x)F_j(x)F_k(x)=\sum\limits_{\substack{{i,j,k}\\{i<j<k}}}d_i(x)d_j(x)d_k(x),\\
\cdots~~~~~~~~~~~\cdots ~~~~~~~~~~\cdots\\
F_1F_2 \cdots F_m=d_1d_2 \cdots d_m\end{array}\end{equation}\vspace{1mm}\\
   Therefore by Lemma 1 and relations (12), we find that \vspace{2mm}\\$T_1\cap T_2\cap \cdots \cap T_m = \langle F_1F_2\cdots F_m\rangle = \langle d_1d_2\cdots d_m\rangle,$\vspace{1mm}\\
		  $T_1+ T_2+\cdots+ T_m=\big\langle \sum\limits_{i=1}^m~d_i-\underset{i<j}{\sum}d_id_j+\underset{i<j<k}{\sum}d_id_jd_k-\cdots (-1)^{m-1}\prod\limits_{i=1}^md_i\big\rangle$,\vspace{1mm}\\ $T_{t_1}+ T_{t_2}+\cdots +T_{t_r}= \big\langle \sum\limits_{i=1}^r~d_{t_i}-\underset{t_i<t_j}{\sum}d_{t_i}d_{t_j}+\underset{t_i<t_j<t_k}{\sum}d_{t_i}d_{t_j}d_{t_k}-\cdots (-1)^{r-1}\prod\limits_{i=1}^rd_{t_i}\big\rangle$.\vspace{1mm}\\ By Proposition 1 (viii) and (ix), we get (i) and (ii).\vspace{2mm}\\
To prove (iii), from Proposition 1 (vi) we see that \vspace{2mm}\\ $E_1E_2= \theta_{r_1}(e_1e_m)+ \theta_{r_2}(e_2e_1)+\cdots +\theta_{r_m}(e_me_{m-1})\\~~~~~~~~=\theta_{r_1}(e_1e_2 \cdots e_m)+ \theta_{r_2}(e_1e_2 \cdots e_m)+\cdots +\theta_{r_m}(e_1e_2 \cdots e_m)=e_1e_2 \cdots e_m$. \vspace{2mm} \\Similarly $E_{t_1}E_{t_2}\cdots E_{t_r}= e_1e_2 \cdots e_m$ for any tuple $(t_1, t_2, \cdots, t_r)$. Hence\\
$E_{t_1}E_{t_2}\cdots E_{t_r}= E_1E_2\cdots E_m $, so we get (iii) by Lemma 1.\vspace{2mm}\\
Again as $E_1+E_2+ \cdots +E_m=e_1+e_2+\cdots+e_m$ and for any tuple $(t_1, t_2, \cdots, t_r)$,
$E_{t_1}E_{t_2}\cdots E_{t_r}= e_1e_2\cdots e_m $, we see that\vspace{2mm}\\
		 $\begin{array}{ll} P_1+ P_2+\cdots+ P_m=\big\langle \sum\limits_{i=1}^m~E_i-\underset{i<j}{\sum}E_iE_j+\underset{i<j<k}{\sum}E_iE_jE_k-\cdots (-1)^{m-1}\prod\limits_{i=1}^mE_i\big\rangle \vspace{1mm}\\
		=\big\langle \sum\limits_{i=1}^m~e_i-\binom{m}{2}e_1e_2 \cdots e_m+\binom{m}{3}e_1e_2 \cdots e_m- \cdots(-1)^{m-1}\binom{m}{m}e_1e_2 \cdots e_m\big\rangle\\=\langle e_1+e_2+ \cdots + e_m - (m-1)e_1e_2 \cdots e_m\rangle.\end{array}$ \vspace{2mm}\\ Now (iv) follows from Proposition 1 (vii).
\vspace{1mm}\\
		  Since $e_j (\overline{j}(x))=0$ for all $1 \leq j \leq m$ by Proposition 2(xv), we get $E_i (\overline{j}(x))=0$ and so $P_i\cap \langle \overline{j}(x)\rangle = \{0\}$. As $d_i=1-e_i$, we find that $F_i=1-E_i$ and so $ F_i (\overline{j}(x))= \overline{j}(x)-E_i (\overline{j}(x))= \overline{j}(x)$. Therefore $T_i\cap \langle \overline{j}(x)\rangle = \langle\overline{j}(x)\rangle$. This proves (v).\vspace{2mm}\\ We prove (vi) for $i=1$. Others are similar. Note that $E_1F_1 = \theta_{r_1}(e_1d_1)+ \theta_{r_2}(e_2d_2)+\cdots +\theta_{r_m}(e_md_m)=0$ and $E_1+F_1 = \theta_{r_1}(e_1+d_1)+ \theta_{r_2}(e_2+d_2)+\cdots +\theta_{r_m}(e_m+d_m)=1$, by Proposition 1(x). Therefore $P_1\cap T_1=\langle E_1F_1 \rangle =\{0\}$ and $P_1+ T_1=\langle E_1+F_1- E_1F_1 \rangle =\langle 1\rangle$.
	 ~~~~~~~~~~~~~~~~~~~~~~~~~~~~~~~~~~~~~~~~~~~~~~~~~~~~~~~~~~~~~~~~~~~$\Box$\vspace{2mm}\\
	\noindent	Similarly we have \vspace{2mm} \\\noindent{\bf Theorem 6:} For any subset $\{t_1, t_2, \cdots, t_r\} \subseteq A,$ where $2\leq r \leq m$,
		the following assertions hold for polyadic codes over $\mathcal{R}$.
		\vspace{2mm}\\
$\begin{array}{ll}
		
		{\rm (i)}& T'_1\cap T'_2\cap \cdots \cap T'_m=  T'_{t_1}\cap T'_{t_2} \cap\cdots \cap T'_{t_r},\vspace{2mm}\\
		{\rm (ii)}& T'_1+T'_2+\cdots+T'_m=\langle 1\rangle= \mathcal{R}[x]/\langle x^n-1\rangle,  \vspace{2mm}\\
		{\rm (iii)}& P'_1\cap P'_2\cap \cdots \cap P'_m= \{0\} ,\vspace{2mm}\\
		{\rm (iv)}& P'_1+ P'_2+\cdots+P'_m= P'_{t_1}+ P'_{t_2}+\cdots+P'_{t_m}, \vspace{2mm}\\
		{\rm (v)} &P'_i\cap \langle \overline{j}(x)\rangle = \{0\}, ~T'_i\cap \langle \overline{j}(x)\rangle = \langle\overline{j}(x)\rangle,\vspace{2mm}\\
{\rm (vi)}& P'_i+ T'_i = \mathcal{R}[x]/\langle x^n-1\rangle, ~P'_i\cap T'_i = \{0\},\vspace{2mm}\\
{\rm (vii)} & P_i+ \langle \overline{j}(x)\rangle =T'_i,~P'_i+ \langle \overline{j}(x)\rangle =T_i, \vspace{2mm}\\
{\rm (viii)}& P_i+P'_i = \langle 1- \overline{j}(x)\rangle, ~P_i\cap P'_i = \{0\} {\rm~and} \vspace{2mm}\\
{\rm (ix)}& T_i+T'_i = \mathcal{R}[x]/\langle x^n-1\rangle,  ~T_i\cap T'_i = \langle  \overline{j}(x)\rangle.\end{array}$
\vspace{2mm}\\

	\noindent \textbf{Proof:}	The proof of statements (i) to (vi) is similar to that of (i) to (vi) of Theorem 5. To prove (vii) we note that \vspace{2mm}
	
 $\begin{array}{ll} E_1+ \overline{j}(x)- E_1 (\overline{j}(x))= E_1+ \overline{j}(x)\\~~~~~~~= \theta_{r_1}e_1+ \theta_{r_2}e_2+\cdots+\theta_{r_m}e_m + \overline{j}(x)(\theta_{r_1}+ \theta_{r_2}+\cdots+\theta_{r_m})\\~~~~~~~= \theta_{r_1}(e_1+(\overline{j}(x))+ \theta_{r_2}(e_2+(\overline{j}(x))+\cdots +\theta_{r_m}(e_m+(\overline{j}(x))\\~~~~~~~= \theta_{r_1}d'_1+ \theta_{r_2}d'_2+\cdots +\theta_{r_m}d'_m= F'_1,~~
		{\rm ~by~ Proposition ~2(xv)}.\end{array}$ \vspace{2mm}

  \noindent Hence  $P_1+ \langle  \overline{j}(x)\rangle =T'_1$. Similarly others. Statements (viii) and (ix) follow from Proposition 2 (xvi).
	 ~~~~~~~~~~~~~~~~~~~~~~~~~~~~~~~~~~~~~~~~~$\Box$\vspace{2mm}

		\noindent{\bf Theorem 7:} Let $P_i$, $P_i'$, for $i=1,2,\cdots,m$, be two pairs of even-like polyadic codes over the ring  with $T_i$, $T_i'$  the associated pairs of odd-like polyadic codes. Then \vspace{2mm}

$~~~~~~~~$ $P_i^{\perp}=\mu_{-1}(T_i)$ and  $P_i'^{\perp}=\mu_{-1}(T_i')$. \vspace{2mm}

\noindent Further  if $\mu_{-1}(e_i)=e_i$ for $i=1,2,\cdots,m$,  then \vspace{2mm}

$~~~~~~~~$ $P_i^{\perp}=T_i$,  $P_i'^{\perp}=T_i'$ and $P_i$, $P'_i$, $T_i$, $T'_i$  are LCD codes over $\mathcal{R}$.
\vspace{2mm}
	
\noindent \textbf{Proof:} By Proposition 1(x), $e_i+d_i=1$. So $\mu_{-1}(e_i)+\mu_{-1}(d_i)=\mu_{-1}(1)=1$. Therefore \vspace{2mm}\\
$\begin{array}{ll} 1-\mu_{-1}(E_1)&=  \theta_{r_1}+ \theta_{r_2}+\cdots+\theta_{r_m}- \mu_{-1}( \theta_{r_1}e_1+ \theta_{r_2}e_2+\cdots+\theta_{r_m}e_m)
\\&=  \theta_{r_1}(1-\mu_{-1}(e_1))+ \theta_{r_2}(1-\mu_{-1}(e_2))+\cdots\theta_{r_m}(1-\mu_{-1}(e_m))\\&= \theta_{r_1}\mu_{-1}(d_1)+ \theta_{r_2}\mu_{-1}(d_2)+\cdots+\theta_{r_m}\mu_{-1}(d_m)\\&=\mu_{-1}( \theta_{r_1}d_1+ \theta_{r_2}d_2+\cdots+\theta_{r_m}d_m)= \mu_{-1}(F_1).\end{array} $\vspace{2mm}\\
Hence $P_1^{\perp}= \langle 1-\mu_{-1}(E_1)\rangle = \langle \mu_{-1}(F_1)\rangle = \mu_{-1}(\langle F_1\rangle)= \mu_{-1}(T_1) $. Similarly, we get the others.\\ Further   if $\mu_{-1}(e_i)=e_i$ for $i=1,2,\cdots,m$,  then by Theorem 5(vi) and Theorem 6(vi),
$$ P_i \cap P_i^{\perp}= P_i\cap T_i =\{0\},~ P'_i \cap P_i'^{\perp}= P'_i\cap T'_i =\{0\}$$
proving thereby that $P_i$ and $P'_i$ are LCD codes over $\mathcal{R}$. Similarly one can check that $T_i$ and $T'_i$ are also LCD codes over $\mathcal{R}$.~~~~~~~~~~~~~~~~~~~~~~~~~~~~~~~~$\Box$\vspace{2mm}

\noindent{\bf Theorem 8:} If  $S_{\infty}'$ is empty, then we have the following additional results:\vspace{2mm}\\
$\begin{array}{ll}

{\rm (i)} & |P_i|= q^{\frac{(\ell k)(n-1)}{m}}, |T'_i|= q^{\frac{(\ell k)(n+m-1)}{m}}. \vspace{2mm}\\
{\rm (ii)} & |P'_i|= q^{\frac{(\ell k)(n-1)(m-1)}{m}}, |T_i|= q^{\frac{(\ell k)(mn-n+1)}{m}}. \end{array}$\vspace{2mm}\\

\noindent \textbf{Proof:} Here since $e_{t_1}e_{t_2}\cdots e_{t_r}=0$, by Proposition 3, we have $E_{t_1}E_{t_2}\cdots E_{t_r}=0$ for ant tuple $(t_1, t_2, \cdots, t_r)$. Therefore for any $s, 1 \leq s \leq m-1, P_1+P_2+\cdots+P_s= \langle E_1+E_2+\cdots +E_s \rangle$ and $(P_1+P_2+\cdots+P_s) \cap P_{s+1}=\langle (E_1+E_2+ \cdots +E_s)E_{s+1}\rangle=\{0\}.$ Hence by proposition 5 (iv),
$$ \begin{array}{ll} | \langle 1- \overline{j}(x)\rangle | = | P_1+P_2+\cdots+P_m|& = \frac{ | P_1+P_2+\cdots+P_{m-1}| | P_m|}{| (P_1+P_2+\cdots +P_{m-1})\cap P_m| } \vspace{2mm} \\& =| P_1+P_2+\cdots+P_{m-1}| | P_m| \vspace{2mm} \\&= \frac{ | P_1+P_2+\cdots+P_{m-2}| | P_{m-1}|}{| (P_1+P_2+\cdots +P_{m-2})\cap P_{m-1}| }|P_m| \vspace{2mm}\\&= | P_1+P_2+\cdots+P_{m-2}||P_{m-1}||P_m|\vspace{2mm} \\& ~~~\cdots ~~~ \cdots ~~~ \cdots ~~~ \cdots \\& =|P_1||P_2|\cdots |P_{m-1}||P_m| = | P_1| ^m. \end{array} $$
As $| \langle 1- \overline{j}(x)\rangle |=(q^{\ell k})^{(n-1)}$, we get that $|P_1|= q^{\frac{(\ell k)(n-1)}{m}}$. Since from Theorems 5 and 6, we have $ P_i+ \langle \overline{j}(x)\rangle =T'_i$ and $P_i\cap \langle \overline{j}(x)\rangle = \{0\}$, $$ |T'_i|=|P_i||\langle \overline{j}(x)\rangle|= q^{\frac{(\ell k)(n-1)}{m}} q^{\ell k} = q^{\frac{(\ell k)(m+n-1)}{m}}.$$ Again from Theorem 6(viii), we see that  $ |P_i||P'_i| = |\langle 1- \overline{j}(x)\rangle | = q^{(\ell k)(n-1)}$, which gives $|P'_i|=  q^{\frac{(\ell k)(n-1)(m-1)}{m}}$. Finally $ P'_i\oplus \langle \overline{j}(x)\rangle =T_i$ gives
$$|T_i|=|P'_i||\langle \overline{j}(x)\rangle|=  q^{\frac{(\ell k)(n-1)(m-1)}{m}} q^{\ell k}=q^{\frac{(\ell k)(mn-n+1)}{m}}.$$

\subsection{Extensions of polyadic codes over the ring $\mathcal{R}$}
When $S_{\infty}'$ is empty, we consider extended polyadic codes over the ring $\mathcal{R}$ which give us some additional results.\vspace{2mm}

	\noindent	Consider the equation \begin{equation} 1+\gamma^{2}n=0. \end{equation} This equation has a solution $\gamma$ in $\mathbb{F}_q$ if and only if $n$ and $-1$ are both squares or both non squares in $\mathbb{F}_q$ (see [11, Chapter 6]).\vspace{2mm}

\noindent For a linear code $C$ of length $n$ over $\mathcal{R}$, $\overline{C}$, the extension of $C$ is defined as \vspace{-4mm}
$$\overline{C} = \{ (c_0,c_1,\cdots,c_{n-1},c_{\infty})~: ~ c_{\infty}= \gamma\sum_{j=0}^{n-1}c_j, ~(c_0,c_1,\cdots,c_{n-1})\in C \}.$$

		\noindent{\bf Theorem 9:} Let $S_{\infty}'$ be empty. Suppose there exists a $\gamma$ in $\mathbb{F}_q$ satisfying equation (13). If the splitting of $\mathbb{Z}_n$ in (1) is given by the multiplier $\mu_{-1}$, then the extended odd-like polyadic codes satisfy $\overline{T_{i+1}'}=\overline{T_i}^{\perp}$.\vspace{2mm}

	\noindent \textbf{Proof:} Here, by Theorem 7, $P_i'^{\perp}=\mu_{-1}(T_i')=T_{i+1}'$. As $T_i=P_i'+\langle \overline{j}(x)\rangle$ and $T_{i+1}'=P_{i+1}+\langle \overline{j}(x)\rangle$, by Theorem 6 (vii), let $\overline{T_i}$ and  $\overline{T_{i+1}'}$ be the extended polyadic code over $\mathcal{R}$ generated by
		$$~~~~~~~\begin{array}{cccccc}
		~~ 0 & 1 & ~2 & \cdots & ~~n-1 & \infty
		\end{array}\vspace{-2mm}$$ $$ \overline{G_i}=\left(
		\begin{array}{ccccc}
		  &  &  &  &~ 0\\
		  &  & G_i' & & ~0 \\
		\vdots &  &  &  & ~ \vdots \\
		 ~1 & 1 &1 ~~\cdots ~~~~&  1 &-n\gamma
		\end{array}
		\right)$$
and $$~~~~~~~\begin{array}{cccccc}
		~~ 0 & 1 & ~2 & \cdots & ~~n-1 & \infty
		\end{array}\vspace{-2mm}$$ $$ \overline{G_{i+1}'}=\left(
		\begin{array}{ccccc}
		  &  &  &  &~ 0\\
		  &  & G_{i+1} & & ~0 \\
		\vdots &  &  &  & ~ \vdots \\
		 ~1 & 1 &1 ~~\cdots ~~~~&  1 &-n\gamma
		\end{array}
		\right)$$
		where $G_i'$ is a generator matrix for the even-like polyadic code $P_i'$ and $G_{i+1}$ is a generator matrix for the even-like polyadic code $P_{i+1}$. The row above the matrix shows the column labeling  by $\mathbb{Z}_n\cup \infty$.  Since the all one vector belongs to $T_{i+1}'$ and  its dual $T_{i+1}'^{\perp}$ is equal to $P_i'$, the last row of $\overline{G_{i+1}'}$ is orthogonal to all rows of $G_i'$. The last row is orthogonal to itself also as $n+\gamma^{2}n^{2} = 0$ in $\mathbb{F}_q$. Therefore all rows of $\overline{G_{i+1}'}$ are orthogonal to all the rows of $\overline{G_i}$. Now the result follows from the fact that $|\overline{T_{i+1}'}|=|\overline{T_i}^{\perp}|$, as can be verified from Theorem 8. $\Box$\vspace{2mm}

\noindent Similarly, we have \vspace{2mm}

	\noindent{\bf Theorem 10:} Let $S_{\infty}'$ be empty. Suppose there exists a $\gamma$ in $\mathbb{F}_q$ satisfying equation (13). If  $\mu_{-1}(e_i)=e_i$ so that the splitting of $\mathbb{Z}_n$ in (1) is not given by the multiplier $\mu_{-1}$, then the extended odd-like polyadic codes satisfy $\overline{T_i}^{\perp}=\overline{T'_i}$.\vspace{2mm}

\noindent{\bf Corollary 1:} If $S_{\infty}'$ is empty, $m=2$ then the following  assertions hold for duadic codes over $\mathcal{R}$.
		\vspace{2mm}\\
		$\begin{array}{ll} {\rm (i)}{\rm ~If~} \mu_{-1}(e_1)=e_2, \mu_{-1}(e_2)=e_1, {\rm~ then}\vspace{2mm} \\
		 ~~~~~ P_i^{\perp} =  T_i', ~ P_i'^{\perp} =  T_i; ~ P_i {\rm ~are~ self~orthogonal} {\rm ~ and ~} \overline{T_i} {\rm ~are~ self~dual}.\vspace{2mm}\\
		{\rm (ii)}  {\rm ~If~} \mu_{-1}(e_1)=e_1, \mu_{-1}(e_2)=e_2, {\rm~ then} ~\overline{T_i} {\rm ~are~ isodual}. \end{array} $\vspace{2mm}\\

\noindent \textbf{Proof:} Here, by definition $\mathbb{C}_1'=\mathbb{C}_2$ and $\mathbb{D}_1'=\mathbb{D}_2$, therefore  $E_1'=E_2$, $F_1'=F_2$, $T_1'=T_2$, $T_2'=T_1$, $P_1'=P_2$ and $P_2'=P_1$.\vspace{2mm}\\
 If $\mu_{-1}(e_1)=e_2$, $\mu_{-1}(e_2)=e_1$, i.e., when the splitting is given by $\mu_{-1}$,  we have by Theorem 7, $P_i^{\perp} =  T_{i+1}$, subscript modulo $m$. Therefore $P_1^{\perp} =  T_2= T_1',  P_2^{\perp} =  T_1=T_2'$.   Using statement (vii) of Theorem 6
, we have $P_i\subseteq T_i'= P_i^{\perp}$. Therefore $P_i$ is self-orthogonal. By Theorem 9, $\overline{T_{i+1}'}=\overline{T_i}^{\perp}$, therefore $\overline{T_i}^{\perp}=\overline{T_i}$.\\
If $\mu_{-1}(e_i)=e_i$, By Theorem 10, $\overline{T_i}^{\perp}=\overline{T'_i}$, therefore $\overline{T_1}^{\perp}=\overline{T_2}$ and $\overline{T_2}^{\perp}=\overline{T_1}.$

\subsection{Griesmer type bound for polyadic codes over $\mathcal{R}$}
 Kuruz et al.\cite{KOS} gave some examples of $m$-adic residue codes over $\mathbb{F}_q[u]/\langle u^2-u\rangle$ whose parameters attain Griesmer type bound. In the next theorem, we prove that the Griesmer type bound for polyadic codes over the ring $\mathcal{R}$ is same as the Griesmer bound for the corresponding polyadic codes over the field $\mathbb{F}_q.$ \vspace{2mm}

\noindent \textbf{Theorem 11:} The parameters of polyadic codes over $\mathcal{R}$  are same as parameters of the corresponding polyadic codes over $\mathbb{F}_q$. Hence  Griesmer type bound for polyadic codes over the ring $\mathcal{R}$ is same as the Griesmer bound for the corresponding polyadic codes over the field $\mathbb{F}_q.$\vspace{2mm}\\
\noindent \textbf{Proof:} Let $\mathcal{C}$ be a polyadic code of length $n$ over $\mathcal{R}={\bigoplus}~\eta_{ij}\mathbb{F}_q.$ Then $\mathcal{C}$ is equal to $T_i$ or  $T_i'$ or $P_i$ or $P_i'$ for $1 \leq i \leq m$. By definition,  $T_1=\theta_{r_1}\mathbb{D}_1 \oplus \theta_{r_2}\mathbb{D}_2 \oplus \cdots + \oplus \theta_{r_m}\mathbb{D}_m $, where $\mathbb{D}_1, \mathbb{D}_2 \cdots \mathbb{D}_m$ are all odd-like polyadic codes over $\mathbb{F}_q$ and are equivalent.  Therefore by Theorem 1,

 $$\begin{array}{ll}k(T_1)&= \max\limits_{i=1}^{m}\{k(\theta_{r_i}\mathbb{D}_i)\} \vspace{2mm}
\\&= {\rm ~ dimension ~of ~polyadic~ code ~} \mathbb{D}_i=k(\mathbb{D}_i)\end{array}$$
$$\begin{array}{ll}~~~~~~~~~~~~~d(T_1)&= \min
\limits_{i=1}^{m}\{d(\theta_{r_i}\mathbb{D}_i)\} \vspace{2mm}
\\&= {\rm ~ minimum~ distance ~of ~polyadic~ code ~} \mathbb{D}_i= d(\mathbb{D}_i).\end{array}$$

\noindent Further $T_i=\mu_{a^{i-1}}(T_1).$ Here Jacobson radical, $J(\eta_{ij}\mathbb{F}_q)=\{0\}$, so $|\eta_{ij}\mathbb{F}_q/J(\eta_{ij}\mathbb{F}_q)|$ $=q$ for every $i$ and $j$. Hence the Griesmer type bound for odd-like polyadic codes $T_i$ over the ring $\mathcal{R}$ becomes $$n \geq {\displaystyle\sum_{i=0}^{k(\mathbb{D}_i)-1} \bigg\lceil \frac{d(\mathbb{D}_i)}{q^i}\bigg\rceil}
,$$
which is same as Griesmer bound for polyadic code $\mathbb{D}_i$ over $\mathbb{F}_q.$\vspace{2mm}\\
Similar result holds for  $T_i'$, $P_i$ and $P_i'.$\vspace{2mm}

\noindent \textbf{Example 1:} Let $q=3$, $n=13$, $m=4$, $f(u)=u^3-u$ and $g(v)=v^2-1$. Take  $E_1=(\eta_{11}+\eta_{12})e_1 + (\eta_{21}+\eta_{22})e_2 + (\eta_{31})e_3 + (\eta_{32})e_4$.
Here $P_1=(\eta_{11}+\eta_{12})\mathbb{C}_1 \oplus (\eta_{21}+\eta_{22})\mathbb{C}_2 \oplus (\eta_{31})\mathbb{C}_3 \oplus (\eta_{32})\mathbb{C}_4$ has parameters $[13,3,9]$. It attains the Griesmer type bound over the ring $\mathbb{F}_{3}[u,v]/\langle u^3-u,v^2-1, uv-vu\rangle.$ Therefore $P_i, ~i=1,2,3,4$ are optimal. The code $T_1=(\eta_{11}+\eta_{12})\mathbb{D}_1 \oplus (\eta_{21}+\eta_{22})\mathbb{D}_2 \oplus (\eta_{31})\mathbb{D}_3 \oplus (\eta_{32})\mathbb{D}_4 $ has parameters $[13,10,3]$. It nearly attains Griesmer type bound over the ring $\mathbb{F}_{3}[u,v]/\langle u^3-u,v^2-1, uv-vu\rangle$.\vspace{2mm}

\noindent \textbf{Example 2:} Let $q=5$, $n=11$, $m=2$, $f(u)=u^3-u$ and $g(v)=v^2-1$. Here $P_1=(\eta_{11}+\eta_{12}+\eta_{21}+\eta_{22})\mathbb{C}_1 \oplus (\eta_{31}+\eta_{32})\mathbb{C}_2 $ has parameters $[11,5,6]$, so it attains the Griesmer type bound over the ring $\mathbb{F}_{5}[u,v]/\langle u^3-u,v^2-1, uv-vu\rangle.$ Therefore $P_1$ and $P_2$ are optimal.  \vspace{2mm}

\noindent \textbf{Remark:} Using the above theory, one can construct some other cyclic codes over the ring $\mathcal{R}$ (which are not polyadic according to our definition) generated by idempotents of the type $\theta_{r_1}(\underset{i \in I_1}{\sum}~e_i)+ \theta_{r_2}(\underset{i \in I_2}{\sum}~e_i)+ \cdots + \theta_{r_m}(\underset{i \in I_m}{\sum}~e_i)$, where $I_1, I_2, \cdots, I_m$ are subsets of $\{1,2,\cdots,m\}$, which may attain the Griesmer type bound. \\For example, take $q=11$, $n=5$, $m=4$, $f(u)=(u^2-1)(u-2)$,  $g(v)=v^2-v$ and $E=(\eta_{11}+\eta_{12}+\eta_{21})(e_1+e_2+e_3)+(\eta_{22}+\eta_{31}+\eta_{32})(e_1+e_2)$. Let $C$ be a cyclic code over ring $\mathcal{R}$ generated by the idempotent $E$, then $C$ has parameters $[5,3,3]$ and it attains the Griesmer type bound.\vspace{2mm}\\
As an another example, take $q=7$, $n=19$, $m=6$, $f(u)=u^4-u$ and $g(v)=v^2-v$ and $E_1=(\eta_{11}+\eta_{12}+\eta_{21})(e_1+e_2)+(\eta_{22}+\eta_{31}+\eta_{32})(e_2+e_3)+(\eta_{41}+\eta_{42})(e_3+e_5)$. The cyclic code $C$ generated by the idempotent $E_1$ over ring $\mathcal{R}$ has parameters $[19,12,6]$ and it nearly attains the Griesmer type bound.

		\subsection{Gray images of polyadic codes over $\mathcal{R}$}

		\noindent \textbf{Theorem 12:} Let the matrix $V$ taken in the definition of the Gray map $\Phi$ satisfy $VV^T=\lambda I_m$, $\lambda \in \mathbb{F}_q^*$. For all possible choices of $(r_1,\cdots,r_m)$, the Gray images of even-like polyadic codes  $ P_i^{(r_1,\cdots,r_m)},  P_i'^{(r_1,\cdots,r_m)}$ and Gray images of extensions of odd-like polyadic codes  $ T_i^{(r_1,\cdots,r_m)},  T_i'^{(r_1,\cdots,r_m)}$, for $i=1,2,\cdots,m $, have the following properties \vspace{2mm}\\
$\begin{array}{ll}(i)& \text{If}~ \mu_{-1}(e_i)=e_i,~ \text{i.e., if}~ \mu_{-1}(S_i)=S_i, ~\text{then} ~\Phi(P_i),~ \Phi(P_i'), ~\Phi(T_i)~ \text{and}~ \Phi(T_i')~ \vspace{1mm} \\& \text{are}  \text{ linear}  \text{ complementary
dual(LCD) codes of length}~ k\ell n ~ \text{over} ~\mathbb{F}_q. \vspace{2mm} \\ (ii) & \text{ If} ~S_{\infty}' ~ \text{is empty and}~ ~ \mu_{-1}(S_i)=S_i, ~ \text{then}~ \Phi(\overline{T'_i})=(\Phi(\overline{T_i}))^{\perp}. \vspace{3mm}\\ (iii) & \text{ If}~ S_{\infty}' ~ \text{is empty and}~ \mu_{-1}(S_i)=S_{i+1}, ~\text{i.e., if the splitting in (1) is given by} \vspace{1mm}\\& \mu_{-1}, ~ \text{then}~ \Phi(\overline{T'_{i+1}})=(\Phi(\overline{T_i}))^{\perp}. \end{array}$\vspace{2mm}

\noindent The theorem follows from Theorems 3, 7, 9 and 10.\vspace{2mm}

\noindent{\bf Corollary 2:} If $S_{\infty}'$ is empty, $m=2$, then the following  assertions hold for duadic codes over $\mathcal{R}$.
		\vspace{2mm}\\
		$\begin{array}{ll} {\rm (i)}&{\rm ~If~} \mu_{-1}(e_1)=e_2, \mu_{-1}(e_2)=e_1, {\rm~ then}~ \Phi(P_i){\rm ~are~ self~orthogonal~ of ~length}~ k\ell n \vspace{2mm} \\&{\rm
and ~}
		  \Phi(\overline{T_i})~\text
{are self-dual codes of length}~ kl(n+1) ~\text{over}~ \mathbb{F}_q.\vspace{2mm}\\
		{\rm (ii)}&  {\rm ~If~} \mu_{-1}(e_1)=e_1, \mu_{-1}(e_2)=e_2, {\rm~ then} ~\Phi(\overline{T_i}) {\rm ~are~ isodual~ codes ~of~ length} \vspace{2mm} \\& k\ell(n+1) ~\text{over}~ \mathbb{F}_q. \end{array} $\vspace{2mm}\\

\noindent The following  examples  illustrate our theory.
	The minimum distances of all these codes have been computed by the Magma Computational Algebra System.\vspace{2mm}

\noindent {\bf Example 3:} Let $q=13$, $n=3$, $m=2$, $f(u)=u^2-u, ~g(v)=v^3-v$, $\gamma=2$ and
  $$V=A =\left(
	\begin{array}{cccccc}
	2 & -2 & 1  & 2 &-2&1\\
	1 & 2 & 2&1&2&2 \\
	2 & 1 & -2& 2&1&-2 \\2&-2&1&-2&2&-1\\1 & 2 & 2&-1&-2&-2\\ 2 & 1 & -2& -2&-1&2

	\end{array}
	\right)\vspace{2mm}$$
be a matrix over $\mathbb{F}_{13}$ satisfying $VV^T=5I$.
 Here  $S_{\infty}'=\emptyset$, $e_1=3x^2+x+9$ and $e_2=x^2+3x+9 $. Also $\mu_{-1}(e_1)=e_2$ and $\mu_{-1}(e_2)=e_1$. On taking $\theta_{r_1}=\eta_{11}+\eta_{12}+\eta_{13}+\eta_{21}+\eta_{22}$ and $\theta_{r_2}=\eta_{23}$, we have $E_1= -x^2(uv^2+uv-3)+x(uv^2+uv+1)+9$ and $ F_1= -x^2(uv^2+uv+1)+x(uv^2+uv-3)+5.$ The Gray images of polyadic codes $P_1^{(5,1)}$ and $\overline{T_1}^{(5,1)}$ are self-orthogonal [18,6,6] and self-dual [24,12,4] codes  over $\mathbb{F}_{13}$ respectively.\vspace{2mm}

\noindent {\bf Example 4:} Let $q=7$, $n=19$, $m=3$, $f(u)=u^2-1,~ g(v)=v^2-v$ and
  $$V=B =\left(
	\begin{array}{cccc}
	2 & -2 & 1  & 1 \\
	-1 & 1 & 2&2 \\
	2 & 2 & 1& -1 \\1&1&-2&2
	\end{array}
	\right)
	$$ be a matrix over $\mathbb{F}_{7}$ satisfying $VV^T=3I$.
The Gray images of polyadic codes $P_1^{(2,1,1)}$, $T_1^{(2,1,1)}$, $P_1'^{(2,1,1)}$ and $T_1'^{(2,1,1)}$ with  $ \theta_{r_1}=\eta_{11}+\eta_{12}, \theta_{r_2}= \eta_{21}$ and $ \theta_{r_3}= \eta_{22}$ are respectively  [76,24,22],  [76,36,12],  [76,52,18] and  [76,28,15] LCD codes  over $\mathbb{F}_{7}$. \vspace{2mm}

\noindent {\bf Example 5:} Let $q=4$, $n=17$, $m=4$, $f(u)=u^2-1,~ g(v)=v^2-v$ and
  $$V=C =\left(
	\begin{array}{cccc}
	a&-a^2&1&1 \\
	-1&1&a&a^2\\
	a^2&a&-1&1 \\1&1&a^2&-a
	\end{array}
	\right)
	$$ be a matrix over $\mathbb{F}_{4}$ satisfying $VV^T=I$, where $a$ is a primitive element of $\mathbb{F}_{4}$.
The Gray images of polyadic codes $P_1^{(1,2,1)}$, $T_1^{(1,2,1)}$, $P_1'^{(1,2,1)}$ and $T_1'^{(1,2,1)}$ with  $ \theta_{r_1}=\eta_{11}, \theta_{r_2}= \eta_{12}+\eta_{21}$ and $ \theta_{r_3}= \eta_{22}$ are respectively  [68,16,28], [68,52,6], [68,48,8] and [68,20,17] LCD codes  over $\mathbb{F}_{4}$.   \vspace{2mm}

\noindent Some other examples are given in Table 1.

\newpage
\begin{center}  \textbf{Table 1.}\end{center} \vspace{2mm}

		{\footnotesize \begin{tabular}{|c|c|c|c|c|c|c|c|c|}
		\hline &&&&&&&& \vspace{-3mm}\\
		$~q$ & $n$ &$m$& $f(u)$ &$g(v)$&$V$&$~\gamma~$&$\Phi({P}_1)$&$\Phi(\overline{T}_1)$ \\ \hline &&&&&&&& \vspace{-2mm}\\
4 & 13 &2&  $u^2-1$& $v^2-u$ & $C$ &1 & [52,24,8] &[56,28,8]\\
		&   &&   & & &   &~LCD~& isodual \\\hline &&&&&&&& \vspace{-2mm}\\
5 & 11 &2&  $u^2-1$& $v^2-1$ & $H_4^{\dag}$ &2 & [44,20,12] & [48,24,11] \\
		&   &&   & & &   & self-orthogonal& self-dual \\\hline &&&&&&&& \vspace{-2mm}\\
		7 & 9 &2&  $u^2-1$&$v^3-v$ & $A$ & ~does not  &[54,24,6] & \\
		&   &  & & &   & exist &~self-orthogonal~~& \\\hline&&&&&&&& \vspace{-2mm}\\
		7 & 3 &2& $u^2-u$& $v^3-v$ & $A$ & 3 &[18,6,6] & [24,12,4]\\
		&   & &  & & &   & self-orthogonal& ~~~~self-dual~~~~~\\\hline &&&&&&&& \vspace{-2mm}\\

        11 & 5 &2  & $u^2-u$& $v^3-v$ & $A$&does not& [30,12,8] & \\
		&   &   &  && & exist  & self-orthogonal &
	\end{tabular}}\vspace{-0.5mm}\\

	{\footnotesize \begin{tabular}{|c|c|c|c|c|c|c|c|c|c|}
	\hline &&&&&&&& \vspace{-3mm}\\

		$q$ & $n$ &$m$& $f(u)$ &$g(v)$&$V$&$\Phi({P}_1)$&$\Phi({T}_1)$&$\Phi({P}'_1)$&$\Phi({T}'_1)$ \\ \hline
3 & 13 &4 & $u^2-1$& $v^2-1$ & $H_4$ & [54,12,24] &[52,40,5] &[52,36,8]& [52,16,13]\\
		&   &   & & &   &&$\sim$ $\Phi({P}_1)^{\perp}$ && $\sim$ $\Phi({P}_1')^{\perp}$\\\hline
5 & 13 & 3& $u^2-1$& $v^2-1$ & $H_4$& [52,16,16] &[52,36,7]& [52,32,8]&[52,20,13]\\
		&   &     & & &
  & LCD & LCD&LCD&LCD\\\hline
		
7$^*$ & 16 &2&  $u^2-1$& $v^2-u$ & $H_4$ & [64,52,2] &[64,12,16]& [64,8,24]& [64,56,2]\\
		&   &  & & &   & LCD&LCD& LCD& LCD \\\hline
11 & 5 &4 & $u^2-u$& $v^2-1$ &$ B$ & [20,4,14] &[20,16,3] &[20,12,5]& [20,8,5]\\
		&   &   & & &   &&$\sim$ $\Phi({P}_1)^{\perp}$ && $\sim$ $\Phi({P}_1')^{\perp}$\\\hline
13 & 17 & 4 & $u^2-1$& $v^2-v$ & $I^{\ddag}$& [68,16,12] &[68,52,4]& [68,48,4]&[68,20,11]\\
&   &  & & &   & LCD&LCD& LCD& LCD \\\hline
16 & 17 & 4 & $u^2-u$& $v^2-v$ & $I$& [68,16,14] &[68,48,5]& [68,52,5]&[68,20,11]\\
&   &  & & &   & LCD&LCD& LCD& LCD \\\hline
32 & 11 & 5 & $u^2-u$& $v^2-v$ & $I$& [44,8,10] &[44,32,4]& [44,36,3]&[44,12,9]\\
&   &  & & &   & LCD&LCD& LCD& LCD \\\hline

	\end{tabular}}\\
{\footnotesize $~~~~$ $^*$In this case, $S_{\infty}'$ is non-empty.\\
$~~~~$ $^{\dag}$$H_4$ is Hadamard matrix of order 4.\\	
$~~~~~$ $^{\ddag}$$I$ is the Identity matrix.}	
\section{Conclusion}
In this paper,  polyadic codes and their extensions  over a finite commutative non-chain ring $\mathbb{F}_{q}[u,v]/\langle f(u),g(v),uv-vu\rangle$ are studied where $f(u)$ and $g(v)$ are two polynomials of degree $k$ and $\ell$ respectively ($k$ and $\ell$ are not both $1$) which split into distinct linear factors over $\mathbb{F}_{q}$. A Gray map is defined from $\mathcal{R}^n \rightarrow \mathbb{F}^{k\ell n}_q$ which preserves  duality. As a consequence,  self-dual, isodual,  self-orthogonal and complementary dual(LCD) codes over $\mathbb{F}_q$ are constructed. Some examples are also given to illustrate our theory. It is shown that the Griesmer type bound for polyadic codes over the ring $\mathcal{R}$ is same as the Griesmer bound for the corresponding polyadic codes over the field $\mathbb{F}_q.$  Examples of some codes which are optimal with respect to Griesmer type bound  are given. The results of this paper can easily be extended over the ring
$\mathbb{F}_{q}[u_1,u_2,\cdots, u_r]/\langle f_1(u_1),f_2(u_2),\cdots f_r(u_r), u_iu_j-u_ju_i\rangle$
 where polynomials $f_i(u_i)$,   $1\le i\leq r$,  split into distinct linear factors over $\mathbb{F}_{q}$.\vspace{2mm}

 \noindent {\bf Acknowledgements}: The research of first author is supported by   University Grants Commission, grant number 405261. The research of second author is supported by Council of Scientific and Industrial Research, CSIR sanction no. 21(1042)/17/EMR-II.

	\end{document}